\RequirePackage{fix-cm}
\documentclass[smallextended]{svjour3}       % onecolumn (second format)
\smartqed  % flush right qed marks, e.g. at end of proof
\usepackage{cite}
\usepackage{amsmath,amssymb,amsfonts}
\usepackage{verbatim}
\usepackage{algorithmic}
\usepackage[ruled,vlined,linesnumbered]{algorithm2e}
\usepackage{graphicx}
\usepackage{multirow}
\usepackage{textcomp}
\usepackage[dvipsnames]{xcolor}
\usepackage{soul}
\usepackage{boldline}
\usepackage{comment}
\usepackage{appendix}
\usepackage{tcolorbox}
\begin{document}

\title{Evaluating Pre-Trained Models for User Feedback Analysis in Software Engineering: A Study on Classification of App-Reviews
\thanks{This research is supported by a grant from Natural Sciences and Engineering Research Council of Canada RGPIN-2019-05175.
% General acknowledgments should be placed at the end of the article.
}
}
% \subtitle{Do you have a subtitle?\\ If so, write it here}

\titlerunning{PTM for App Review classification}        % if too long for running head

\author{Mohammad A Hadi \and Fatemeh H Fard %etc.
}


\institute{M. A. Hadi \at
              Dept. of Computer Science \\
              University of British Columbia\\
              Kelowna, Canada\\
              Tel.: +1 250 899 6971\\
              \email{mohammad.hadi@ubc.ca}           %  \\
           \and
           F. H. Fard \at
              Dept. of Computer Science \\
              University of British Columbia\\
              Kelowna, Canada\\
              \email{fatemeh.fard@ubc.ca} 
}

\date{Received: date / Accepted: date}

\maketitle

\begin{abstract}
    
    \textit{Context:} 
    %What is your research about? Why are you doing this research, why is it interesting?
    Reviews of mobile apps on App Stores and on social media are valuable resources for app developers. Analyzing app reviews have proved to be useful for many areas of software engineering (e.g., requirement engineering, testing, etc.). 
    Existing approaches rely on manual curating of a labeled dataset to classify app reviews automatically. In practice, new datasets must be labeled as the classification purpose changes (e.g., identifying bugs versus usability issues or sentiment), limiting the power of developed models to classify new classes or issues.
    Employing models that can produce the same or better result using less labeled data can eliminate the problem by reducing this manual effort.
    Recent Pre-trained Transformer based Models (PTM) are trained on large natural language corpora in an unsupervised manner to retain contextual knowledge about the language and have found success in solving similar Natural Language Processing (NLP) problems. However, the applicability of PTMs has not yet been explored for issue classification from app reviews. 
    
    \textit{Objective:}
    %What exactly are you studying/investigating/evaluating? What are the objects of the study? We welcome both confirmatory and exploratory types of studies.
    We explore the advantages of PTMs for app review classification tasks by comparing them with existing models; we also examine the transferability of PTMs by applying them in multiple settings. 

    \textit{Method:}
    %How are you addressing your objective? What data sources are you using?
    We select six datasets from the literature which contain manually labeled app review from Google Play Store, Apple App Store, and Twitter data.
    We empirically study the performance and time efficiency of PTMs compared to Prior approaches.
    In addition, we evaluate the performance of the PTMs when app reviews are incorporated in their pre-training (i.e., domain-specific PTMs). 
    We set up different studies to evaluate PTMs in multiple settings: binary vs. multi-class classification, zero-shot classification (when new labels are introduced to the model), multi-task setting, and classification of reviews from different resources. 
    In all cases, Micro and Macro Precision, Recall, and F1 scores are used and the time required for training and prediction with the models are also reported. 
    
    \textit{Results:}
    Our results show that PTMs can classify the app issue with higher scores. However, in the multi-resource setting, the Ensemble method and Deep learning with word embedding score higher than PTMs. When PTMs are pre-trained with app specific reviews (i.e. Custom PTMs), the models achieve the highest scores among all in all settings, and the prediction times are also reduced; with the highest scores belonging to customized PTMs that are pre-trained with a larger number of app reviews. 
    
    \textit{Conclusion:}
    For app issue classification, the PTMs are the best choices in some settings, compared to Prior approaches; but they are not the best ones for all the studied settings. It is worth considering the customized PTMs pre-trained on app reviews when higher performance and lower prediction times are required. 
    
\end{abstract}

\section{Introduction}
\label{sec:intro}

    %Give more details on the bigger picture of your study and how it contributes to this bigger picture. An important component of phase 1 review is assessing the importance and relevance of the study questions, so be sure to explain this.
    
    Mobile application (app) marketplaces, such as Apple App Store and Google Play enable the users to rate and review apps \cite{hadi}. 
    Users express their usage experience through writing reviews from different perspectives, such as the app's quality, performance, and functionality. 
    The reviews provide a {distinct} way for the application developers to acquire customer feedback \cite{luliang}. 
    %The app developers use these comments for requirement engineering and release planning \textcolor{red}{[REF]} to update the app.
    The developers monitor the reviews regularly to address users' major concerns and resolve the reported issues in the forthcoming app update.   
    If the app is not regularly updated addressing the issues, it will gradually lose its popularity \cite{liu2015, islam2010}. 
    %As the mobile app users specifically write about their experience on App Stores revealing myriads of information about the app to the software engineers, the developers monitor the reviews regularly. 
    Therefore, app reviews have been studied extensively \cite{2020survey} and are shown to be resourceful for requirement engineering \cite{luliang}, release planning \cite{ciurumelea2017analyzing}, software maintenance  \cite{al2020classification}, change-file localization \cite{zhou2020TSE}, and testing \cite{grano2018testing}. 
    %to increase app ranking, which eventually can affect the app profit, finding issues, requirement engineering, release planning, and software maintenance. 
    %Identifying the key topics for developers, finding emerging topics for the released version of an app, 
    Different studies also focused on analyzing the reviews to filter non-informative reviews \cite{arminer}, identify bugs \cite{maalej}, classify usability and quality concerns \cite{luliang}, and find security and privacy issues \cite{besmer2020privacy}. 
    
    Other than App Stores, mobile app users also express their opinions on social media such as Twitter, which regularly reveal new information for the developers \cite{stanik}. %Even though it is easier for app developers to get useful information from this review channel than traditional media (such as bug/change repositories, crash reporting systems, online forums, emails, etc.), some unique obstacles prevent them from doing so. 
    Nonetheless, the manual extraction of informative feedback from App Stores or social media is difficult, as there are many unrelated and noisy user comments. 
    Therefore, researchers have developed techniques, including rule-based, machine learning, or deep learning approaches, for automatic extraction of useful information from user reviews to help app developers \cite{arminer, gukim, guzman, stanik}. 
    There are many studies that classify the reviews for different purposes, which range from identifying problems, user inquiries, feature requests, and aspect evaluations (e.g., feature strengths, weaknesses, and performance) to sieving usability, portability, and reliability \cite{stanik, 2020survey, luliang, ciurumelea2017analyzing, maalej, besmer2020privacy}. 
    These studies provide app developers with a variety of serviceable information that facilitates making informed decisions for planning the app updates. 
    This extra information is crucial for app developers as mobile applications and their updates get released quite frequently over a short period to meet the market requirements; so, it often proves challenging to identify and prune \textit{all} the software defects and bugs during the testing phase \cite{Joorabchi}.
    
    The topics of interest in app {review} classification are numerous and different studies leverage various supervised machine learning techniques to extract useful software engineering information from app reviews \cite{2020survey}.
    A main requirement of supervised techniques is the availability of labeled datasets, which needs to be done manually {by field-experts \cite{stanik}}.
    Although app review analysis is shown to be useful for software developers in practice, this manual effort is expensive and seems to be a barrier in many studies \cite{2020survey}.
    Recent works try to alleviate the problem by incorporating semi-supervised learning \cite{deocadez2017semi-supervised} or active learning \cite{dhinakaran2018active-learning}.
    However, with the emergence of new labels of interest, e.g., privacy or security issues \cite{besmer2020privacy}, developers must label new datasets, which renders the previously curated datasets obsolete. 
    In addition, the distribution of the data collected from different platforms (i.e., App Stores or social media) varies. Therefore, even though the same labels are to be classified from two platforms, a model that has been trained on one (e.g., Google Play) needs to be retrained to extract the same labels from another (e.g., Twitter) to yield similar performance \cite{stanik}. 
    %\textcolor{blue}{Moreover, to mine and classify Software Engineering (SE) related issues from app-reviews, the developed supervised-issue classifiers need to simultaneously learn the lexical and semantic properties of natural language as well as the inherent qualities from SE-related texts. As these tools/methods do not have any prior knowledge of Natural Language (NL), they demand large manually curated datasets, which are labor-intensive [REF].} 
    Whenever the classification purpose (i.e., labels of interest) or platform changes, the models require retraining on newly labeled datasets since they do not retain the natural language properties learned from previous training, including the linguistic features and the context of the trained labels \cite{maalej}.
    {Training from scratch prevents the extensibility of the developed models for classifying new issues in practice.}
    
    A more recently established and widely accepted practice in Natural Language Processing (NLP) is using Pre-Trained Language Models (PTM) and then transferring its learned knowledge to various downstream NLP tasks, such as sentiment analysis, question answering, or classification \cite{qiu2020pre}. 
    In NLP, PTMs (such as BERT) are large language models that are trained on large natural language corpora using a deep neural network in an unsupervised manner \cite{how-far-ptm}. 
    These models are then fine-tuned for various downstream tasks using limited labeled datasets. 
    %For example, BERT \cite{bert} is a PTM that is frequently being used for question answering and sentiment classification tasks.
    As PTMs are trained on large general domain corpora, they learn contextual linguistic information and eliminate the need to train downstream task models from scratch \cite{wada2020medicalBERT}. PTMs reduce the amount of 
    {effort (i.e., new model development time per task)} to build models for each task separately, and they reduce the amount of required labeled dataset \cite{roberta}.
    Consequently, PTMs are used to transfer the learned knowledge to a new domain or a new task, and in settings where a model has not seen any example of the required task or interested label during training (known as zero-shot learning) \cite{zero}. 
    Although PTMs are used extensively and led to many advances in NLP, their applicability for software engineering is barely known. 
    Only a few studies exist that explore the use of PTMs for sentiment analysis in software engineering \cite{how-far-ptm, biswas2020achieving}, or for tasks related to programming languages such as comment generation \cite{Svyatkovskiy}.
    But, to what extent PTMs can be applied for app review classification is unknown.

    %\textcolor{blue}{, which show that traditional machine learning models can exceed PTMs' performance on selected datasets \cite{how-far-ptm, biswas2020achieving}.}
    Moreover, the PTMs that are trained on domain-specific data, show significant improvements over general purpose PTMs such as BERT (i.e. PTMs that are trained on general purpose data) for NLP tasks in fields such as science or law \cite{sci-bert, legal-bert}.
    These domain-specific PTMs leverage unsupervised training on a large domain-specific corpus to compensate for the lack of high-quality, large-scale labeled data in the specified domains.  
    Our problem of interest, the classification of app reviews, aims to extract useful information for software engineers/app developers, which are domain-specific information. In addition, app reviews are short text, and many reviews are noisy and unrelated, which can introduce more complications \cite{hadi}. To what extent the PTMs can be useful in the context of app review classification and whether domain-specific PTMs trained on app reviews can improve the results is unknown.
    Previously, the deep learning approaches used Convolutional Neural Networks (CNN) to implement the transfer of static linguistic knowledge from non-contextual word embedding, which has shown to have comparable results with traditional machine learning approaches for classification of requirements related app reviews \cite{stanik}. In contrast, another study using CNN shows improvements over previous approaches for app review classification \cite{aslam2020convolutional}. Though these studies do not use PTMs, they use a deep learning approach which sheds doubt on whether PTMs can be useful.

    Therefore, in this study, we aim to explore the benefits of PTMs compared to the existing approaches for app review analysis, specifically the app review classification tasks.
    We define the app review classification as the task of extracting useful information from users' feedback which can be related to app requirements, release planning, and software maintenance. The extracted information may help identify different aspects of the application, such as feature requests, aspect evaluations (e.g., feature strength, feature shortcoming, application performance), usability, portability, reliability, energy consumption, problem reports, privacy, and security inquiries about the application. 
    Our goal is to investigate the accuracy and time efficiency of PTMs for the app issue classification task over different selected datasets from literature with various labels and multiple tasks (i.e., issue classification and sentiment analysis of app reviews).
    Therefore, experiments will be conducted in different settings: by fine-tuning the PTMs on different sizes of the labeled dataset for downstream tasks, exploring the same PTMs for multiple tasks in app review analysis, and finally, comparing the performance of PTMs when they are trained on non-domain-specific dataset versus PTMs trained on the domain-specific dataset.
    These experiments will provide baselines on the applicability of PTMs for app review analysis, including the cost of using them (in terms of required time for predictions) and their capability to reduce the manual effort required to label large datasets. 
    %\textcolor{blue}{Our results show that the PTMs can achieve at least the same performance as the current approaches and be beneficial to be used for multiple classification tasks. }
    
    The contributions of this study are as follows:
    % \vspace{-1mm}
    \begin{itemize}
    	\item This is the first study that explores the applicability of PTMs for automatic app issue classification tasks compared to the existing tools.
    	\item We conduct an extensive comparison between four PTMs and four existing tools/approaches on six different app review datasets with different sizes and labels.
    	\item We are the first to explore the performance of general versus domain-specific pre-trained PTMs for app review classification.
    	\item This is the first empirical study to examine the accuracy and efficiency of PTMs in four different settings: binary vs. multi-class classification, zero-shot setting, multi-task setting, and setting in which training data is from one resource (e.g., App Store) and the model is tested on data from another platform (e.g., Twitter).
    \end{itemize}

We explore the following research questions for our study:

\textbf{RQ1: How accurate and efficient are the PTMs in the classification of app reviews compared to the existing tools? }
%established tools and approaches (i.e., rule based tools, feature engineered ML algorithm, or DL model)?
In this research question, we explore how the current PTMs perform compared to the existing tools, including their required time for training and prediction. The existing tools are based on curated rules, feature engineered machine learning algorithms, and deep learning models. 
%The results can give us insights about the applicability of using PTMs in practice. 

\textit{Findings:}
\textcolor{black}{We find that different PTMs outperform Prior approaches by $\sim$3\% to $\sim$15\% on all datasets. Additionally, the best-performing PTMs require slightly more time for prediction, although the difference in prediction time is negligible (+0.04s to +2.87s).}

\textbf{RQ2: How does the performance of the PTMs change when they are pre-trained on app-review dataset, instead of a generic dataset (e.g. Wiki-documents, book corpus)?} %, Google News)?
The current PTMs are trained using general text scraped from the web. In some studies, domain-specific models are trained e.g. LEGAL-BERT \cite{legal-bert} and medical BERT \cite{wada2020medicalBERT} which have been shown to improve the performance compared to the non-domain-specific models. In RQ2, we intend to explore the performance of a domain-specific pre-trained model when trained on app reviews. 
%The results will help the researchers focus on building domain specific models (e.g. building app review models in languages other than English) versus using the general PTMs. 

\textit{Findings:}
\textcolor{black}{The PTMs trained from scratch on domain-specific data (i.e. adding app-reviews to the model) performed better than out-of-the-box models (+0.8\% to +2.2\% micro-F1 score). The Custom PTMs trained from scratch do not fluctuate much in prediction times with respect to the readily available PTMs for app issue classification. In addition, we find that incorporating a greater number of app reviews in the pre-training can help PTMs produce up to 15.2\% better micro-F1 score.}

\textbf{RQ3: How do the PTMs perform in the following settings? \\
    (RQ3-1) Binary vs multi-class setting, \\
    (RQ3-2) Zero-shot classification, \\
    (RQ3-3) Multi-task setting (i.e. different app-review analysis tasks),  \\ %(e.g., issue classification, sentiment analysis), and \\
    (RQ3-4) Classification of user-reviews collected from different resources (i.e., Twitter,  App Store).}

The answers to this research question will help us understand the applicability of PTMs and their performance in different settings (i), when the classification is on two or more classes; (ii), when a labeled dataset is not available; (iii), the task changes; and (iv) different resources are used. In part (iv) we explore the transferability of the models as the distributions of data on various platforms is different \cite{ruder2017transfer}. 
%We expect that PTMs achieve reasonable results for these settings.

\textit{Findings:}
\textcolor{black}{Both Prior and PTM models yield better performance for binary
classification tasks than for multi-class settings. PTMs are the best choice for zero-shot classification settings. Custom PTMs improve the results of their non-domain-specific models. For multi-task and multi-resource settings, the Custom LARGE PTMs (i.e., PMTs pre-trained with 10 million app reviews) performed better than readily available PTMs. 
RoBERTa-based Custom PTMs have the best scores among all models. They also have lower prediction times in the multi-resource setting than Prior approaches.}

\textbf{Deviations from the Registered Report.}
The following are the parts that we have changed compared to the Registered Report.
\begin{itemize}
    \item In the multi-task setting (RQ3-3), we add experiments on another dataset, in addition to the previous dataset mentioned in the Registered Report. This new dataset is on app reviews, but classifies them as different categories (e.g. game, family). This new dataset is added to assess the performance of these models for a different classification task but in the same domain.
    \item Another difference of this submission with the original plan in the Registered Report is on the multi-resource setting (RQ3-4). Here, we add experiments on the Prior approaches as well to examine their capabilities for this setting and compare the results of PTMs against them.
\end{itemize}

\textbf{Paper Organization. }
The rest of the paper is organized as follows.
We review the related works in Section \ref{sec:relWork}.
Section \ref{sec:study} discusses the Study Overview, including research design, datasets, selection of the approaches, and experimental setup. In Section \ref{sec:approach_rq} we detail our approach to answer each of the research questions and discuss the results in Section \ref{sec:results}. Section \ref{sec:disc} is dedicated to the discussions. Threats to validity are in Section \ref{sec:threats} and we conclude the paper in Section \ref{sec:conclusion}.

%% \section{Background Knowledge}
%% \label{sec:back_knowledge}
%% 
%% \textcolor{red}{TODO: remove or change to related work? Let's think more!
%% Talk about what are PTMs in general, the Transformer architecture very briefly, how the PTMs are used (fine-tuned), and ... }

\section{Related Works}
\label{sec:relWork}

\subsection{\textbf{App Review Classification in Software Engineering}}

\subsubsection{\textbf{Topic Modeling Based Models}}

According to Silva et al. \cite{silva2021topic}, topic modeling has been applied to software engineering research, including app review classification. Among the most frequent topic modeling techniques used to categorize app reviews, the researchers found that Latent Dirichlet Allocation (LDA) \textcolor{black}{\cite{lda}} and LDA-based techniques are the most common.
AR-Miner is one of the initial works for mining app reviews proposed by Chen et al. \cite{arminer}, which uses topic modeling to group the informative reviews and investigates categories of users' discussions.
Nayebi et al. \cite{nayebi2018app} examined how Twitter app reviews can contribute to mobile app development and applied topic modeling and crowdsourcing. %, they analyzed an additional 22.4\% of feature requests and 12.89\% of bug reports from Twitter and concluded that these data sources ultimately provide developers with additional value.
Adaptive Online LDA was developed by \textcolor{black}{Gao et al. to classify app reviews based on users' feedback on various versions of mobile apps \cite{gao2018online}.}
Consequently, Hadi and Fard developed Adaptive Online Biterm Topic Model (AOBTM) \cite{hadi2020aobtm} to model topics of app reviews in different categories adaptively, therefore, alleviating the sparsity problems in short-texts, which considered the statistical data for multiple previous time slices.

Topic modeling approach is used in other works \cite{gao2018online, SRM-citer2, yang2021tour, wardhana2021aspect, gao2018online} to categorize app reviews for detecting emerging issues for developers to update their apps, identifying fine-grained app features in the reviews to extract users' sentiments, prioritizing important user reviews for developers, and performing aspect-based sentiment analysis of app reviews.

\subsubsection{\textbf{Machine Learning Based Approaches}}

Earlier research has focused on using machine learning approaches for app review filtering. Chen et al. \cite{arminer} classified non-informative reviews by training a classifier and categorizing them into informative and non-informative groups. Using a regression model, Fu et al. \cite{making_sense} filter out rating-inconsistent reviews that differ in sentiment from their ratings. In \cite{gukim}, Gu et al. identified software features and distinguish between features that are evaluated and those that are requested, using Max-Entropy \textcolor{black}{\cite{nigam1999maxEnt}}. There are other studies that used machine learning techniques for app issue classification: \cite{shah2018simple} and \cite{Rustam2020class} use lexical features and multi-text features to use in conjunction with different machine learning models for app review classification task. \textcolor{black}{Maalej and Nabil} \cite{maalej} empirically evaluated different feature extraction techniques to work with machine learning models to automatically classify bug reports and feature requests. Martin et al. \cite{EM-citer1} and Sarro et al. \cite{sarro2015} have also explored the breadth of research done on App Store analysis (i.e., feature life cycle, dormant feature identification) for software engineering.

In \cite{guzman, Triantafyllou2020how, Messaoud2019multi, fu2021machine, guo2019enhancing}, researchers have used ensemble machine learning techniques to classify app reviews into categories relevant to software evolution, to identify more efficiently and earlier terms in reviews that could be classified into specific topics, and to filter out useful feedback from users, such as feature requests, bug descriptions, and requirement analysis.

\subsubsection{\textbf{Deep Learning Based Approaches}}

Using machine learning, \cite{stanik} demonstrated that they could still achieve comparable results with deep learning when analyzing user feedback in English and Italian into problem reports, inquiries, and irrelevant.
\textcolor{black}{Aslam et al.} proposed an approach for app reviews classification using non-textual information to classify app reviews and exploiting deep learning technique \cite{aslam2020convolutional}.
In \cite{mekala2021class}, a BERT-based sequence classifier was developed and validated to achieve a state-of-the-art average classification accuracy (87\%) for feedback analysis. %This approach proved effective even in extremely low-volume dataset environments, drastically reducing the time and costs of evaluation and improving accuracy measures.
From a large volume of online product reviews, \textcolor{black}{Qiao et al.} \cite{qiao2020deep} proposed to apply a domain-oriented approach to deep learning to discover the most critical users' needs, such as app product new features and bug reports.
\textcolor{black}{Henao et al.} \cite{henao2021transfer} investigated the prospect of transfer learning for the classification of app reviews and found that monolingual BERT models outperform existing baseline methods in the classification of English App Reviews.

Other studies \cite{he2019detecting, haering2021automatic, Aralikatte2018fault, wang2020analyzing, hadi, zhao2019sentiment} also applied deep learning techniques to different extents, such as detecting promotion attacks, matching bug reports, detecting adversarial spam, analyzing energy-related reviews, and requirement evolution predictions from app reviews.

\subsection{\textbf{Pre-trained Models}}

There are two generations of pre-trained models: static and contextual word embeddings \cite{ptm_survey}. The Global Vectors for Word Representation (Glove) \cite{glove} and word2vec \cite{w2v} are examples of first-generation static word embeddings. Software Engineering (SE) research has used word embeddings for various tasks, including code retrieval \cite{code_retrieval}, detecting incoherent comments \cite{comment_coherence}, specifying SE-relevant tweets \cite{sieve}, and program comprehension with graphs \cite{program_representation}. These kinds of embeddings are context-independent, where each word has only one embedding and cannot change with different contexts. These embeddings only take the syntax of the words in a sentence into account. Therefore, non-contextual embeddings cannot capture semantic information and model polysemous words \cite{ptm_survey}.

The second-generation pre-trained models learn context-sensitive word representations and can be tailored to perform downstream tasks effectively. LSTM \cite{lstm}-based Universal Language Model Fine-tuning (ULMFiT) \cite{ulmfit} and Embeddings from Language Models (ELMo) \cite{deep_context} have also been used for different SE-related tasks, such as identifying ambiguous software requirements \cite{ulmfit_citer} and sentiment analysis for software engineering \cite{elmo_citer}. Transformer architectures \cite{attention} have been used to build several pre-trained Transformer-based models \cite{bert, albert, roberta, xlnet} that have achieved state-of-the-art performance for a variety of downstream tasks, such as sentiment classification of SE-related texts \cite{how-far-ptm} and identifying error-prone software \cite{error_prone}.

Other studies \cite{Rajpurkar2016sqaud, Reddy2019cpqa, yang2018hotpotqa, Zhang_retrospective, ju_technical, tu_select} have used PTMs for solving Question-Answering problem. PTMs have also been extensively used the sentiment analysis task \cite{Bataa_investigation, sun2019bert, xu2019bert, Rietzler_adapt, karimi_adversarial, li2019exploiting, wux_mask, peters2017semi}. \textcolor{black}{Previous studies} \cite{liu2018efficient, hakala2019biomedical, edunov2019ptm} developed Named Entity Recognition systems with the help of PTMs. Others \cite{clinchant2019bert, imamura2019recycling, zhang2019hibert} studied how PTMs can improve machine translation tasks. Other works include using PTMs for text summarization \cite{liu2019text, zhong2020extractive, jin2020BERT} and adverserial defense and attacks identification \cite{wallace2019universal, sun_adv, lil_bert, zhuc2020freelb, liu_adversarial}.

Though the studies that use PTMs or classify app reviews are numerous, there is no previous work that investigates their capability for app issue classification as we conduct in this work.

\section{Study Overview}
\label{sec:study}

In this section, we first overview the design of the research, and then provide details about the selected datasets, Prior approaches, and chosen pre-trained models. We also discuss what metrics will be used to evaluate the models, as well as explain the experimental setup. %the details about model training and experimental setup.

\subsection{Research Overview}

\begin{figure}[hbt!]
    \centerline{\includegraphics[width=0.7\linewidth]{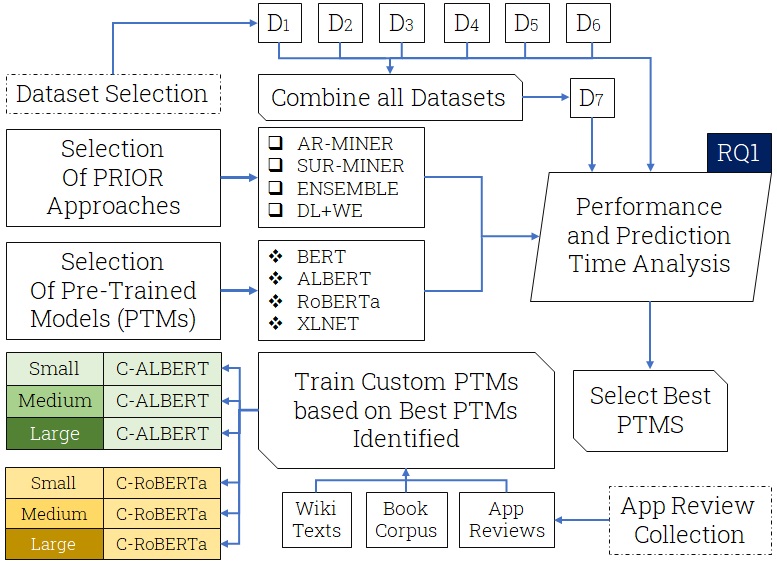}}
    \caption{Overview of Research Steps for RQ1}
    \label{fig:intro2}
\end{figure}

\begin{figure}[hbt!]
    \centerline{\includegraphics[width=1.0\linewidth]{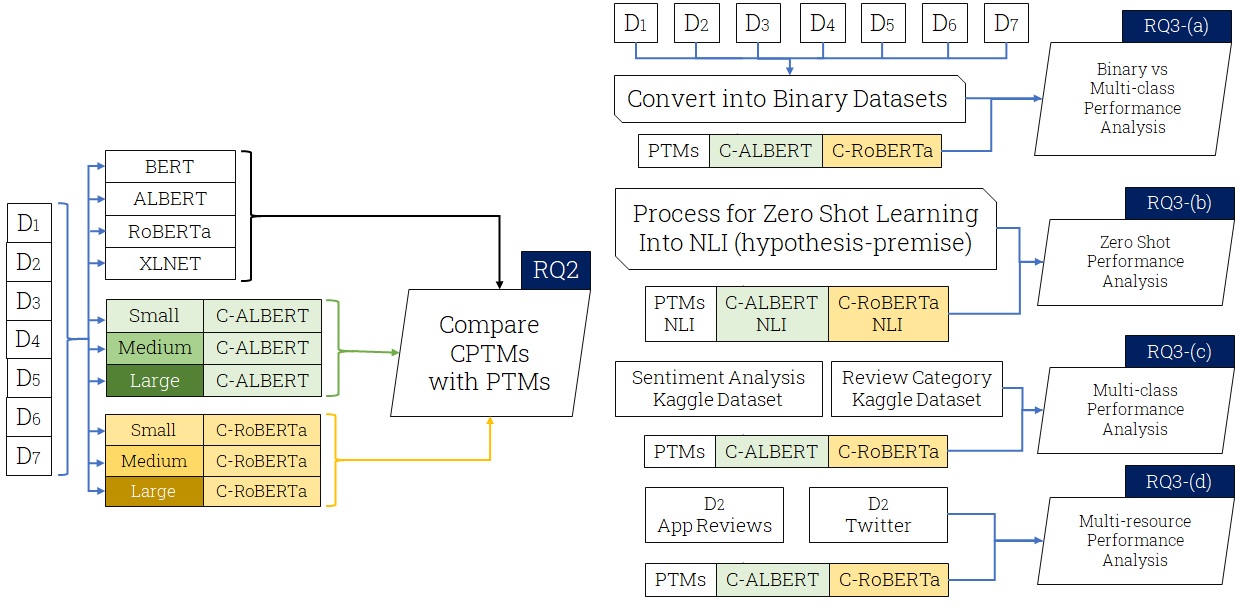}}
    \caption{Overview of Research Steps for RQ2 (left) and RQ3 (right)}
    \label{fig:intro3}
\end{figure}

\textcolor{black}{Figures \ref{fig:intro2} and \ref{fig:intro3} show the overall process of our study. First, we select 6 widely used datasets from the literature for app review classification. We also combine these datasets and make another dataset, which has multiple labels and is highly imbalanced ($D_7$).
We choose four approaches from the literature that we refer to as Prior approaches. 
Four pre-trained language models are selected to evaluate their capabilities regarding the research questions in this study.
We evaluate all the Prior approaches and the PTMs for the datasets. From the PTMs, the best performing ones are selected to train Custom PTMs (\texttt{CPTM}) and assess them in RQ2 for all the seven datasets. 
In RQ3, we evaluate the performance of the PTMs and Custom PTMs in four different settings. The details of the datasets and selected approaches and PTMs are explained in next sub-sections.}

\subsection{\textbf{Datasets}}

App reviews are analyzed automatically for various purposes and reviews are classified with different labels such as user concerns, feature requests, feature modifications, bug reports, and usability analysis {\cite{aobtm}}. 
In our experiments, we examine the performance of PTMs for app issue classification. 
Therefore, we select a diverse set of datasets from the literature as described below. 
These datasets have various sizes, different labels, and collected from different platforms (i.e., Google Play, Apple App Store, Twitter), and therefore, can represent to what extent the PTMs can be useful for app issue classification. 

\textbf{Dataset 1 ($D_1$)}
This dataset is procured by Gu and Kim \cite{gukim} and contains reviews for 17 popular Android apps from Google Play in 16 most popular categories, such as games and social. The authors have manually labeled 2,000 reviews for each app (34,000 in total) in five classes according to their predefined rules: Aspect Evaluation (5,937), Praise (8,112), Feature Request (2,323), Bug Report (2,338), and Others (15,290).
  
\textbf{Dataset 2 ($D_2$)}
This dataset contains 6,406 app reviews from Google Play and 10,364 tweets (sampled from 5 million tweets written in English) which are labeled manually into three classes: Problem Report, Inquiry, and Irrelevant \cite{stanik}.
They included Problem Report (1,437), Inquiry (1,100), and Irrelevant (3,869) records from Google Play. The number of records in each of these categories from Twitter data is 2,933, 1,405 and 6,026, respectively.

\textbf{Dataset 3 ($D_3$)}
This dataset is provided by Lu and Liang. \cite{luliang}. Researchers selected two popular Apps, one from Apple App Store (iBooks in the books category) and one from Google Play (WhatsApp in the communication category). 
They sampled the raw reviews collected from each platform and manually classified 2,000 reviews for each app (4,000 in total). 
The reviews are classified into six categories: 
Usability (432), Reliability (587), Portability (119), Performance (121), Feature Request (558), and Others (2,183).

\textbf{Dataset 4 ($D_4$)}
This dataset was procured by Maalej and Nabil \cite{maalej} and contains 2,000 manually labeled reviews from random apps selected from top apps in different categories (1,000 from Apple App Store and 1,000 from Google Play), where half of the apps are paid and half of them are free apps. 
The authors ensured that there are 200 reviews for each of the 1, 2, 3, 4, and 5 stars in each of the 1000-reviews datasets.
In addition, they provide 2,400 more reviews which are manually labeled. 
These reviews are for selected 3 random Android apps and 3 iOS apps from the top 100 apps (400 reviews for each app). 
This dataset contains 4,400 labeled reviews in total with four categories: Bug Report (378), Feature Request (299), User Experience (737), and Rating (2721).

\textbf{Dataset 5 ($D_5$)}
This dataset is published by Guo et al. \cite{guo2020caspar} and contains 1,500 app reviews, which are manually labeled as User Action (428), App Problem (399), and Neither (673).
The reviews are selected from over 5.8 million records for 151 apps from Apple App Store. % from 2008 until 2017.  

\textbf{Dataset 6 ($D_6$)}
This dataset was procured by Guzman et al. \cite{guzman}; it contains reviews of 3 apps from Apple App Store and 4 apps from Google Play. The apps are popular and from diverse categories.
They sampled 260 reviews per app and five annotators labeled the records, summing to 1,820 reviews that are manually labeled. 
This dataset includes seven categories, namely: Bug Report (990), Feature Strength (644), Feature Shortcoming (1281), User request (404), Praise (1703), Complaint (277), and Usage Scenario (593)\footnote{Note that adding numbers in all categories will exceed the total number, because some reviews belong to multiple groups. We will follow the steps in \cite{guzman} to calculate the evaluation metrics for this dataset.}.

\vspace{3mm}
\textit{Reasons behind choosing these Datasets ($D_1$ to $D_6$):}
We considered these datasets for their diverse characteristics from three perspectives, other than having different categories. 
\textit{First}, the considered datasets were constructed from various repositories: 
Apple App Store, Google Play, and Twitter.
\textit{Second}, the size of the datasets vary and they have different number of records for each label. 
For example, \textit{Dataset 2} is around four times larger than \textit{Dataset 5}. Also, \textit{Dataset 2} contains approximately 2,000 reviews per label, whereas \textit{Dataset 1} has around 400 reviews per label. The size differences can provide insights about the ability of PTMs for different sizes (total training set and the available data for each label). 
\textit{Third}, except \textit{Dataset 1} and \textit{Dataset 5} which are balanced dataset, the other four datasets are imbalanced. 
These differences ensure that we explore the capability of PTMs for different sizes, platforms, and in more realistic settings. 
    % A fourth and a trivial aspect is that \textit{Dataset 1, 6} provided the app-wise labels, where others did not. 

\vspace{3mm}
\textbf{Merged Dataset ($D_7$)} % and Compiled Test Dataset (CTD)}}
    \textcolor{black}{This dataset will be compiled by merging all the 6 datasets, $D_1$--$D_6$. This \textit{Merged Dataset} has 55,933 app reviews with 16 labels. As some of the labels used in the datasets are the same or have similar definitions, we have grouped some of them.
    %For example, Aspect Evaluation from $D_1$ includes aspects of the app that are both about Feature Strengths and Shortcomings, similar to $D_6$. Therefore, we grouped them together.
    For grouping the labels, we have consulted the definitions of the classes from the studies that published the datasets $D_1$--$D_6$. %, and also confirmed the final labels with two software engineering researchers with ten years of experience who are not the authors of this paper. 
    %This resulted in 7 classes in $D_7$.
    This dataset would be closer to practical applications, where a lot of irrelevant data exists and multiple classes of informative reviews are of interest to be extracted from this highly imbalanced dataset. 
    %We will be able to clarify the generalizability of our study by using this dataset.
    The labels in the Merged Dataset, the grouped labels from $D_1$--$D_6$, and the number of app reviews per group are provided in Table \ref{table:annotation}}.

    \begin{table}[htbp]
    \caption{Description for Merged Dataset (D7)}
    
    \begin{center}
    \renewcommand{\arraystretch}{1.2}
    % \resizebox{1.0\linewidth}{!}{%
    \begin{tabular}{m{2cm} | m{4.5cm} | m{1.5cm} }
        \hline
        {\textbf{Labels in Merged Dataset}} & {\textbf{Grouped Labels [Label Name: Dataset Initial (\# of Reviews)]}} & {\textbf{\# App Reviews in Total}}\\
        \hline
        {Performance} & {Performance: D3 (121)} & 121\\
        \hline
        {Portability} & {Portability: D3 (119)} & 119\\
        \hline
        {Usability} & {Usability: D3 (432)} & 432\\
        \hline
        {Reliability} & {Reliability: D3 (587)} & 587\\
        \hline
        {Usage Scenario} & {Usage Scenario: D6 (593)} & 593\\
        \hline
        {Feature Strength} & {Feature Strength: D6 (644)} & 644\\
        \hline
        {User Experience} & {User Experience: D4 (737)} & 737\\
        \hline
        {Feature Shortcoming} & {Feature Shortcoming: D6 (1,281)} & 1,281\\
        \hline
        {Inquiry} & {Inquiry: D2 (1,100), User Action: D5 (428)} & 1,528\\
        \hline
        {Problem} & {Problem Report: D2 (1,437), App Problem: D5 (399), Complaint: D6 (277)} & 2,113\\
        \hline
        {Rating} & {Rating: D4 (2,721)} & 2,721\\
        \hline
        {Bug Report} & {Bug Report: D1 (2,338), Bug Report: D4 (378), Bug Report: D6 (990)} & 3,706\\
        \hline
        {Feature Request} & {Feature Request: D1  (2,323),  Feature Request: D3 (558), Feature Request: D4 (299), User request: D6 (404)} & 3,584\\
        \hline
        {Aspect Evaluation} & {Aspect  Evaluation: D1 (5,937)} & 5,937\\
        \hline
        {Praise} & {Praise: D1 (8,112), Praise: D6 (1703)} & 9,815\\
        \hline
        {Irrelevant} & {Others: D1 (15,290), Irrelevant: D2 (3,869), Others: D3 (2,183), Neither: D5 (673)} & 22,015\\
        \hline
        {\textbf{Total}} & {} & {\textbf{55,933}} \\
        \hline
    \end{tabular}
    % }
    \end{center}
    \label{table:annotation}
    \end{table}

%In contrast to the original individual datasets, the \textit{Merged Dataset} better replicates the real-life practical data, which is often more imbalanced and may consist of a variety of classes. We will use the \textit{Merged Dataset} to fine-tune and train the PTMs and Prior approaches, respectively.

%    \textcolor{red}{We have manually labeled 350 app reviews to compile a new dataset, where equal number (50) of reviews are available for each label in \textit{Merged Dataset}. We denote this dataset as \textit{Compiled Test Dataset (CTD)}. After training or fine-tuning on MD, reproducing similar results on CTD would improve the external validity of our study.}

\subsection{\textbf{Prior approaches}} % (PRIOR)}

In this section, we discuss the four widely-used approaches for app review analysis and the reasons of choosing them in our experiments. 
\textcolor{black}{We have conducted literature review and reviewed analysis of app review surveys to select these Prior approaches.}

%We refer to these approaches collectively as the PRIOR group.

\textbf{AR-Miner}
Chen et al. proposed App Review Miner (AR-Miner) \cite{arminer} which extracts valuable information from user reviews. 
%AR-Miner filters the non-informative reviews using data mining and ranking techniques. 
Provided a collection of user reviews, AR-Miner first applies a pre-trained classifier that separates non-informative reviews. Then, AR-Miner applies Latent Dirichlet Allocation (LDA) over the informative reviews to chunk them into different groups for prioritizing them by an efficient ranking model proposed by the authors.
%Finally, AR-Miner visualize the ranking results in a concise and intuitive way to help app developers spot the key feedback users have.
AR-Miner has been widely used to mine app-issues from user-reviews \cite{panichella2015can, ARM-citer2, ARM-citer3, ARM-citer4, ARM-citer5, ARM-citer6}. Therefore, it is considered as a prior approach in our experiments. 

\textit{Reason behind choosing AR-Miner:} 
%\textcolor{blue}
{AR-Miner is an unsupervised approach and its performance could be evaluated against other models. We mainly choose these techniques to evaluate the PTMs against unsupervised, supervised, ensemble methods using classical machine learning approaches, and finally deep learning technique that is enriched by contextual embedding.}
\textcolor{black}{AR-Miner is the first framework to accommodate application developers in mining informative topics from a large volume of app reviews. This was the first effective attempt to leverage an unsupervised Topic Modeling technique to extract edifying issues from app reviews. Developers embraced AR-Miner for its ability to filter out non-informative reviews and display the key user feedbacks in an intuitive, concise manner.  
The effectiveness of AR-Miner was validated by conducting a comprehensive set of experiments on user reviews of four Android apps. 
Researchers compared the AR-Miner results against real app developers’ decisions in different studies \cite{panichella2015can, ARM-citer2, ARM-citer3, ARM-citer4, ARM-citer5, ARM-citer6}. They analyzed the advantages of AR-Miner over manual inspection and other techniques used in a traditional channel \cite{finkelstein, johann, moran, sarro2015}. Based on the empirical results, AR-Miner has proven to be effective and efficient in extracting informative reviews.}

\vspace{3mm}    
\textbf{SUR-Miner}
Gu et al. \cite{gukim} proposed Software User Review Miner (SUR-Miner), which is a framework that summarizes users’ sentiments, opinions, and emotions toward different aspects of an application. 
SUR-Miner 
parses aspect-opinion pairs from reviews by considering their structures. For this purpose, it uses pre-defined sentence templates/patterns\footnote{https://guxd.github.io/srminer/appendix.html}. Then, for each review, it combines the sentiment of the sentences with its aspect-opinion pairs. These are used in the final step to summarize the software aspects. 
SUR-Miner generates reliable summaries and achieved high F1 score for aspect-opinion identification, sentiment analysis, and app review classification compared to the previous works \cite{SRM-citer1, SRM-citer2} {by using a simple but effective machine learning algorithm, \textit{Max-Entropy}}. It is adopted in many studies \cite{SUR-citer3, palomba2017recommending, SUR-citer4, SUR-citer5, features} and therefore we choose it as a prior approach.

\textit{Reason behind choosing SUR-Miner:} \textcolor{black}{SUR-Miner is the first framework that fully took advantage of user reviews’ monotonic structure and semantics; it also defined sentence patterns to extract aspect opinion pairs from app-review. Researchers carefully selected five distinct text features, specifically tailored for app reviews, and leveraged these features to train a supervised machine learning classifier. Based on the empirical evaluation, we choose this machine learning approach as the terminal classes categorized by the SUR-Miner model were shown to be significantly more accurate and precise than other methods, demonstrating its effectiveness \cite{SRM-citer1, SRM-citer2}.}
    
\vspace{3mm}
\textbf{Ensemble Methods}
Guzman et al. systematically evaluated different sets of ensemble methods and identified one for classifying user reviews \cite{guzman}. 
They selected machine learning techniques that were used for text classification, and compared the performance of each algorithm individually, for app review classification. 
They also studied the performance of these models when their results are combined using ensemble methods. 
Ensemble methods are well known machine learning techniques as they can enhance the prediction performance of single classifiers, maintaining their strengths and reduce their vulnerabilities. 
In this work, Logistic Regression, Naive Bayes, Support Vector Machines, and Neural Network Classifier were grouped to vote for the final prediction. 
This ensemble method outperformed the individual algorithms with statistical significance. 
%\textbf{\textit{Reason behind choosing this Ensemble Method:}}  Researchers have engineered customized features to work with specific traditional ML model, but these engineered features do not warrant consistency over different applications and ML algorithms \cite{maalejea}. 
It was tested on a large dataset of seven diverse apps, where it shows that in all cases, this ensemble method either outperformed or matched its best baseline. Also, this method influenced further research in the field of app review analysis \cite{EM-citer1, EM-citer2, ciurumelea2017analyzing, EM-citer4, EM-citer5, EM-citer6} and therefore is used in our study.

\textit{Reason behind choosing Ensemble Methods:} \textcolor{black}{Different studies have attempted to identify the best machine learning approach for app issue classification without spending too much time perfecting custom-engineered features for different issues \cite{maalejea, augmenting, luliang, changelog, features}. Guzman et al. \cite{guzman} first leveraged different machine learning algorithms and combined their predictions to circumvent the problem of various machine learning models being used for classification of issues from app reviews. Their study showed that recall (an evaluation metric) improved significantly after combining the predictions of Logistic Regression and Neural Networks. Software developers and software evolution experts widely adopted this approach to analyze user reviews and prioritize their tasks \cite{feedback, alsubaihin, categorizing, findingFeatures}.}
    
\vspace{3mm}
\textbf{Deep Learning Model leveraging Non-contextual Word Embedding}
Stanik et al. used a Deep Convolutional Neural Network (CNN) \cite{stanik} for classifying app reviews. Their model contains an embedding layer and its weights are initialized with a word embedding model, e.g. word2vec or FastText. 
This CNN architecture outperforms the shallow Neural Networks classifying app reviews in the determined groups, with a small margin.
This method uses non-contextual word embedding within a deep neural network and performed a transfer of knowledge representation. This approach is one of the first studies leveraging transfer learning and deep learning and is used in other studies for issue classification task \cite{mcnn-citer1, mcnn-citer2, aslam2020convolutional}, and therefore, we choose it as a prior approach in our study.

\textit{Reason behind choosing this Deep Learning Approach:} \textcolor{black}{This study is a recent work with the purpose of understanding and evaluating the extent to which deep learning models could be used to categorize user feedback into different predefined categories \cite{stanik}. Based on this study, researchers determined that the domain experts’ knowledge incorporation with traditional machine learning model could achieve comparable results to those of the deep learning approach due to the substantial performance improvements provided by using simple yet powerful features in the traditional machine learning techniques. Nevertheless, the developers and researchers widely adopted this deep learning approach as the improvement of pre-trained word embeddings bring considerable performance gain to the classifier in other domains \cite{yang2018, reimers2019, ren2016}. }

It is worth noting that the number of studies for app review classification is numerous. We choose these four Prior approaches as they are well cited and widely used in many other studies. More importantly, they use different machine learning and deep learning approaches which can be representative of multiple techniques used in other studies. 
\textcolor{black}{Additionally, these models are open sourced and are publicly available.} The best of these four approaches in terms of the performance metrics defined in Section \ref{sec:evaluation-metrics} will be chosen as the baseline.

% If the reviewers find them appropriate, we will consider adding more Prior approaches.
%\footnote{
We have already considered two more tools that we eliminated from our study: AR-Doc and CLAP. 
AR-doc uses a parser based on a NLP Classifier, which uses a set of 500 sentence-structures manually evaluated for four types of classes: Information Giving, Information Seeking, Feature Request, and Problem Discovery. To use AR-Doc, we will need to build different sets of sentence structures for each dataset and their respective labels, which is infeasible. 
In addition, we could not find any publicly available implementation of CLAP and therefore unable to use CLAP\footnote{We contacted the authors of CLAP but did not receive a response.}.
%}.   
    
\subsection{\textbf{Pre-Trained Models (PTM)}}

In this section, we briefly discuss our choice of the four PTMs. All these models have Transformer deep learning architecture which is based on attention mechanism \cite{xlnet}. 
As Transformer is currently the main architecture for PTMs \cite{roberta}, we choose the following Transformer PTMs: BERT, XLNet, RoBERTa, and ALBERT. 
%We refer to these pre-trained Transformer based models collectively as ''PTM". 
The same PTMs are also used for sentiment classification in software engineering \cite{how-far-ptm}. 
    
\vspace{3mm}
\textbf{BERT}
 %(Bidirectional Encoder Representations from Transformers)
Devlin et al. \cite{bert} designed Bidirectional Encoder Representations from Transformers (BERT) to learn contextual word representations from unlabeled texts. Contextual word embeddings designate a word's representation based on its context by capturing applications of words across different contexts.
BERT employed a bidirectional encoder to learn the words' contextual representations by optimizing for Masked Language Model (MLM) and Next Sentence Prediction (NSP) tasks.
For MLM, 15\% of all the tokens are replaced with a masked token (i.e., [MASK]) beforehand, and the model is trained to predict the masked words, based on the context provided by the non-masked words.
For NSP, the model takes sentence-pairs as input for learning to predict whether a pair-match is correct or wrong. During training, 50\% of the inputs are true consequent pairs, while the other 50\% are randomized non-consequent sentence-pairs.
Devlin et al. trained two versions: small-sized BERT\textsubscript{BASE} and big-sized BERT\textsubscript{LARGE}. BERT\textsubscript{BASE} is a smaller model with 12 layers and 110 million parameters. BERT\textsubscript{LARGE} has 24 layers and 340 million parameters.
BERT\textsubscript{LARGE} is more computationally expensive and consumes more memory compared to BERT\textsubscript{BASE}. Thus, in this work, we will use BERT\textsubscript{BASE}.
    % Please note that based on the results reported in the BERT paper, BERT\textsubscript{LARGE} always exceeds BERT\textsubscript{BASE}.
    
\textit{Reason behind choosing BERT:} BERT advanced the state-of-the-art for 11 NLP tasks. It achieved an absolute improvement of 7.6\% over the previous best score on General Language Understanding Evaluation (GLUE) benchmark\footnote{https://gluebenchmark.com/}; it also achieved 93.2\% accuracy on SQuAD 1.1\footnote{Stanford Question Answering Dataset: https://rajpurkar.github.io/SQuAD-explorer/explore/1.1/dev/}, outperforming human performance.
As the tasks in these datasets require natural language understanding and include classification tasks, we choose BERT.

\vspace{3mm}
\textbf{XLNet}
XLNet \cite{xlnet} uses Auto-Regressive language modeling and Auto-Encoding. 
    %In XLNET, the next token is dependent on all previous tokens. 
The model is “generalized” because it captures the bi-directional context using a mechanism called Permutation Language Modeling (PLM). 
XLNet integrates auto-regressive models and bi-directional context modeling, yet overcoming the disadvantages of BERT.
PLM is the idea of capturing bidirectional context on all permutations of terms present in an input sequence. 
XLNet discards the one-directional linear modeling to maximize the log-likelihood over all permutations of the sequence-terms. Each position is expected to learn utilizing contextual information from the entire sequence, thereby capturing the bidirectional context. No [MASK] is needed, and input data need not be corrupted.

In addition, XLNET addresses capturing the dependency between masked positions, which is neglected by BERT. 
Consider the sentence “New Delhi is a city.” Then, input to BERT to be “[MASK] [MASK] is a city,” and the objective of BERT would be predicting "New" given "is a city" and predicting "Delhi" given "is a city".
In this objective, there is no dependency between learning “New” and “Delhi.” So, BERT can result in a prediction like “New Gotham is a city.”
If we assume that the current permutation is [is, a, city, New, Delhi], BERT would predict the tokens 4 and 5 independent of each other. Whereas, XLNet, predicts in the order of the sequence. i.e., first predicts token 4 and then predicts token 5: it computes 
the likelihood of "New" given "is a city" plus the likelihood of "Delhi" given "New, is a city".

\textit{Reason behind choosing XLNet:} XLnet outperformed BERT on many NLP tasks; for 8 different tasks XLNet beat BERT by a substantial margin. This model achieved best results for 18 NLP tasks, including sentiment classification and natural language inference. As XLNET is outperforming other models on text classification tasks, we choose it as one of the PTMs in our study.

\vspace{3mm}
\textbf{RoBERTa} 
Robustly optimized BERT approach (RoBERTa) outperformed all the state-of-the-art benchmarks upon release \cite{roberta}. Liu et al. modified BERT's pre-training steps that yield substantially better performance on all the classification tasks. 
RoBERTa increased the amount of mini-batch sizes, data, and training time to train the model. 
RoBERTa is also trained on dataset that includes longer sequences than before. The masking pattern in RoBERTa was also modified to be generated spontaneously.
    
\textit{Reason behind choosing RoBERTa:} RoBERTa outperforms BERT on nine different NLP tasks on the GLUE benchmark; it also equals or exceeds XLNet's model in four out of nine individual tasks. Based on these results, RoBERTa can present a reasonable choice for PTM in our study.
    
\vspace{3mm}
\textbf{ALBERT} 
%The first sentence is wrong?
%ALBERT \cite{albert} \textcolor{blue}{has a much larger architecture than BERT, which makes it computationally more expensive}. 
% [FROM ALBERT PAPER] While ALBERT-xxlarge has less parameters than BERT-large and gets significantly better results, it is computationally more expensive due to its larger structure.
A Lite BERT (ALBERT) \cite{albert} applies three parameter reduction techniques: Factorized embedding parameterization, Cross-layer parameter sharing, and Inter-sentence coherence loss. 
In the first one, researchers separated the hidden layers' size from the input embeddings' size (previously of same sizes). They projected one-hot vectors to embedding and the hidden space with lower dimensions; it increased the hidden layer-size without significantly increasing the vocabulary embeddings' parameter size.
Second, all the parameters across all layers are shared. The big-scale ALBERT model has substantially fewer parameters than BERT\textsubscript{LARGE}.
Finally, the NSP task is swapped with Sentence-Order Prediction (SOP) loss which help ALBERT perform better.

\textit{Reason behind choosing ALBERT:} ALBERT uses parameter reduction techniques which resulted in having 18 times less parameters than BERT. Its' training time is 1.7 faster and has negligible inferior performance than the original BERT\textsubscript{LARGE} model.
The much larger ALBERT architecture, which contains fewer parameters than BERT\textsubscript{LARGE}, achieved higher F1 scores 
%of 92.2\% and 89.4\% 
on the SQuAD 2.0 and the GLUE benchmarks; it also achieved high accuracy on the ReAding Comprehension from Examinations benchmark (RACE benchmark).
Therefore, we choose ALBERT as one of the PTMs in our experiments to evaluate its performance and inference time for app issue classification, because ALBERT is shown to be faster than other PTMs without sacrificing much performance. 

\textcolor{black}{For BERT, XLNET, RoBERTa and ALBERT, their base versions are used. Table \ref{table:ptm_config} includes the names of the models used from the Huggingface Transformers library \cite{huggingface} and default configurations.}

\begin{table}[htbp]
\caption{Details of PTMs Used in Our Study}

\begin{center}
% \renewcommand{\arraystretch}{1}
% \resizebox{1.0\linewidth}{!}{%
\begin{tabular}{l | c | c | c | c | c}
\hline
    \textbf{Architecture} & \textbf{Used Model} & \textbf{Parameters} & \textbf{Layers} & \textbf{Hidden} & \textbf{Heads} \\
\hline
    BERT & bert-base-cased & 110M & 12 & 768 & 12 \\
    RoBERTa & roberta-base & 125M & 12 & 768 & 12 \\
    XLNet & xlnet-base-cased & 110M & 12 & 768 & 12 \\
    ALBERT & albert-base-v1 & 11M & 12 & 768 & 12 \\
\hline
\end{tabular}
% }
\end{center}
\label{table:ptm_config}
\end{table}

\subsection{\textbf{Implementation}} 

%\textcolor{red}{TODO: Add details about how each model (prior and PTM) is trained, details about preventing overfitting, details about hyperparamter tuning, etc for DNNs and CPTMs. The details of dataset prep for RQ2 and 3 and more details to be included in the next section?!}
We are going to discuss the implementations of both Prior and PTM approaches for the mentioned datasets.

\subsubsection{\textbf{Prior approaches}}
\textbf{AR-Miner} uses LDA \cite{lda} topic modeling algorithm to categorize text document corpus into a predefined number of different related and coherent topics. For app review classification, we hid the labels of the considered datasets and presented the app reviews altogether as a text corpus, where each review represented different documents. If a dataset has $n$ classes, we would extract $n$ number of topics from the app review corpus; so, the documents will be separated into $n$ number of segments. After the extraction of topics, we looked into the associated labels for each review for assigning each document cluster (topic) to a relevant class/label. The relevance was measured by the prevalence of a class or label in a topic after extraction. Suppose, the highest number of app reviews belong to label/class $x$ in a topic (document cluster), we would understand that AR-Miner has labeled all documents in that cluster with class $x$, where the app reviews that truly belonged to class/label $x$ are regarded as the True Positives and the rest of the reviews are regarded as False Positive.

\textbf{SUR-Miner} used \textit{Max-Entropy} model to classify app reviews. This classifier model considers all probability distributions empirically consistent with the training data and chooses the distribution with the highest entropy.  Empirically consistent probability distribution with training data is one in which an estimated frequency of occurrence of a class and a feature value is equal to the actual frequency in the training data. Following the literature \cite{gukim}, we have implemented Generalized Iterative Scaling (GIS) as our scaling method and performed Early Stopping to avoid over-fitting and to stop the iterative solver from taking a long time to optimize over a large number of feature weights. As mentioned in \cite{gukim}, we have also adopted 2-4 Grams (Character N-Gram) for the classification task.

\textbf{Ensemble Approach} adopted an ensemble of Logistic Regression classifier and shallow Neural Network classifier, and applied the majority voting scheme to combine the output of the classification approaches, by following the literature \cite{guzman}. We have concatenated TF-IDF vector representations of app reviews, number of words in the review, number of
characters in the review, number of lower case characters, number of upper case characters to use as features for our ensemble approach. We have used Adam optimizer and binary cross entropy as our loss function in the shallow neural network.

\textbf{Deep Learning approach with Word Embeddings} followed \cite{stanik}. For the neural network's input layer, text inputs must have a fixed size. This input size has been fixed at 400 words, a measure we found suitable for all the app reviews in all of the datasets we studied, and we identified no app reviews that exceeded 400 words. In addition to the input layer, our network consists of an embedding layer (pre-filled with a fast-text model), a 1D convolution layer, a 1D global max-pooling layer, a dense layer, and a final output layer with a softmax activation. Tanh activation was used for the previous layers. The embedding layer weights were frozen during training, leaving the trainable parameters.

\subsection{\textbf{Experimental Setup}}
\label{experimental-setup}

\textcolor{black}{We will train all the Prior approaches and fine-tune all the PTMs on each of the datasets separately. In the following, we provide the general information about training the models using stratified k-fold cross validation, as well as how PTMs will be trained and a classifier is added on top of each model. }{The detailed process for each research question is provided in section \ref{sec:approach_rq}}. 

\textcolor{black}{To avoid introducing bias to the results due to probable differences in the distribution of the training and test splits,} following \cite{stanik}, we will use k-fold cross validation on each dataset $D_i$ separately, where $D_i$ represents one of the seven datasets in our study. We use k-fold cross validation as a more rigorous approach than splitting the dataset into train and test set once \cite{how-far-ptm}. Also, this method is used previously for app review classification in the software engineering domain \cite{micromacro}.
In the k-fold cross validation, we split a dataset $D$ into $k$ equal size disjoint parts/folds $D\textsuperscript{(1)}, ..., D\textsuperscript{(k)}$. We then build $k$ classifiers $c\textsuperscript{(i)}$, each time using $k-1$ splits as training and one part $D\textsuperscript{(i)}$ as test set. This way, each split of the data can be used as test set once. 
As a result, we will have $k$ different test set performances.
As the datasets in the study are imbalanced, we will use an alternate option of cross validation called stratified k-fold cross validation \cite{apple, imbalancedLearning}. 
In this method, the only difference is that the distribution of the examples from each class in the original dataset is preserved in each fold $D\textsuperscript{(i)}$. 
The stratified k-fold cross validation is commonly used in machine learning practices as it reduces the experimental variance. Therefore, when comparing different methods, it is easier to identify the best method \cite{apple}.
\textcolor{black}{
We consider $k=5$ as this value has been shown empirically to have test error rate estimates that do not suffer from high bias or high variance \cite{james2013introduction}. The value of $k=5$ is also used in previous app review classification studies \cite{stanik}. 
For the evaluation, we use metrics explained in section \ref{sec:evaluation-metrics}.
It is worth mentioning that we apply the stratified 5-fold cross validation on each of the datasets $D_1$--$D_7$ separately. For example, we split $D_1$ into 5 folds $D_1\textsuperscript{(1)}, ..., D_1\textsuperscript{(5)}$ and build 5 classifiers for $D_1$. We then compute the evaluation metrics for $D_1$. We continue this process for each of the datasets $D_2$--$D_7$ and report the performance on each of them separately.}

Note that the number of labels in the datasets $D_1-D_6$ ranges between [3,7] and
all the considered Prior approaches were designed for multi-class classification. 
Therefore, we can retrain \textcolor{black}{and evaluate them using the stratified 5-fold cross validation.} 
Among the Prior approaches, only AR-Miner requires some adaptation. AR-Miner has three phases: classification of non-informative reviews, topic modeling of informative reviews, and ranking of the groups according to their relevance to developers' applications. As all datasets $D_1$--$D_7$ are labeled, 
we will directly utilize the second step to group all the reviews (from both training and test set) in a dataset into the corresponding number of labels available. Following the literature \cite{arminer}, we will count the training set reviews in a resulting group and use the label with the majority count to annotate the group. 
This annotation indicates the class that the reviews in the group belong to. Then, we will examine the group's test set reviews to compute the evaluation metrics.

For evaluating the PTMs, for each dataset, we will use the training set to fine-tune the PTMs and then evaluate them on the test set. We will obtain all the four PTMs that are pre-trained on general domain corpora from Hugging Face library\footnote{https://huggingface.co/}.  
To build a classifier, a feed-forward dense layer and Soft-max activation function will be added on top of each model. 
%The pre-trained model’s parameters can be reused as a starting point. 
%For each dataset, we feed the training set to a pre-trained model’s tokenizer and then train the PTM to get the respective fine-tuned model. Finally, we will test it on the test set.
\textcolor{black}{We follow \cite{how-far-ptm} and \cite{bert} to set the hyper-parameter values. In our work, we set the batch size to 16 and Adam learning rate to $2e-5$. The models are trained for 4 epochs and \textit{AdamW} optimizer is used for all models.}

\textcolor{black}{We execute all the experiments on a Linux machine with Intel 2.21 GHz CPU and 16GB memory. For training PTMs from scratch, we use 2 $\times$ NVIDIA Tesla V100 32GB to enhance the parallelization performance.}

%\textcolor{blue}{In contrast, SUR-Miner utilizes a supervised machine learning approach (Max Entropy) to categorize reviews based on distinct features. To make the datasets compatible with this approach, we will need these features extracted from all the datasets.}

\subsection{\textbf{Evaluation Metrics}} 
\label{sec:evaluation-metrics}

In this section, we describe the evaluation metrics. For all the classifications, we will use three metrics: Precision (P), Recall (R), F1 score (F1), and their micro and macro average values. 
To answer the time efficiency of the models, we will examine the training time for the Prior approaches and the fine-tuning time for PTMs. We will also report the prediction times and changes in time (increase or decrease of time compared to a baseline).

\textbf{\textit{Precision (P):}} Precision can be calculated by dividing the number of records that their labels are correctly predicted by total number of predicted observations in that class: $
%\begin{equation}
P = \frac{TP}{TP+FP}.
%\end{equation} 
$
Here, TP refers to the number of records that their label is correctly predicted, and FP refers to the the number of records falsely predicted to belong to this class. 
In multi-class classification, for each group \textit{A},
all the observations that belong to other labels and are falsely predicted as group \textit{A} are added to compute the FP. 

\textbf{\textit{Recall (R):}} For each group \textit{A}, Recall can be calculated by dividing the number of accurately predicted observations in \textit{A} by the number of all observations available in the corresponding class: $
%\begin{equation}
R = \frac{TP}{TP+FN}
%\end{equation}
$
Here, FN is the number of observations in class \textit{A} which are falsely predicted as other labels.

\textbf{\textit{F1 Score (F1):}} F1 score is the weighted average of Precision and Recall:

\begin{equation}
\label{eq:f1}
%\resizebox{0.25\hsize}{!}{$
    \mbox{F1} = \frac{2 \cdot (P \cdot R)}{P + R}
%    $}
\end{equation}

As we have multi-class classification and the datasets are imbalanced, we use micro-average and macro-average metrics as follows. Here, the micro-average calculates the contribution of all records in all classes (therefore the contribution of the class with the predominant number of records is taken into account), whereas the macro-average is the average of the values for each class (therefore, each class contributes equally in the final value). 
The micro- and macro-averaged precision (P) are computed as:

\begin{equation}
%\resizebox{0.5\hsize}{!}{$
        P_{micro} = \frac{\sum_{j=1}^m TP_j}{\sum_{j=1}^m TP_j + \sum_{j=1}^m FP_j}
%    $}
\end{equation}
\begin{equation}
%\resizebox{0.33\hsize}{!}{$
        P_{macro} = \frac{\sum_{j=1}^m P_j}{m}
%   $}
\end{equation}

\vspace{-1mm}
$TP_j$ and $FP_j$ are the number of true positive and false positive predictions for the $j$-th class, respectively. $P_j$ is the precision for class $j$ and $m$ is the number of classes. 
Similarly, the micro- and macro-average of Recall (R) and F1 score will be calculated and denoted as R\textsubscript{micro}, R\textsubscript{macro}, F1\textsubscript{micro}, and F1\textsubscript{macro}.

\textcolor{black}{
As we are using stratified 5-fold cross validation, for each dataset $D$, we will report averages of these metrics obtained from each of the k classifiers: 
\begin{equation}
%\resizebox{0.5\hsize}{!}{$
    F1\textsuperscript{avg}\textsubscript{micro} = 1/k \sum_{n=1}^{k} F1\textsubscript{micro}\textsuperscript{(i)}
%   $}
\end{equation}
\begin{equation}
%\resizebox{0.5\hsize}{!}{$
    F1\textsuperscript{avg}\textsubscript{macro} = 1/k \sum_{n=1}^{k} F1\textsubscript{macro}\textsuperscript{(i)}
%   $}
\end{equation}}

\textcolor{black}{The $F1\textsubscript{micro}\textsuperscript{(i)}$ and $F1\textsubscript{macro}\textsuperscript{(i)}$ refer to the test performance of classifier $c\textsuperscript{(i)}$ on held out test set $D\textsuperscript{(i)}$. %, considering that we train k classifiers $c\textsuperscript{(i)}$ and we evaluate each classifier on the held out test set $D\textsuperscript{(i)}$. 
Similarly, the P\textsuperscript{avg}\textsubscript{micro}, P\textsuperscript{avg}\textsubscript{macro}, R\textsuperscript{avg}\textsubscript{micro}, and  R\textsuperscript{avg}\textsubscript{macro} will be computed.}

Following previous works \cite{how-far-ptm}, \textit{if a model has higher values for both F1\textsuperscript{avg}\textsubscript{micro} and F1\textsuperscript{avg}\textsubscript{macro}, we consider it to be better than other models.}

% \textcolor{red}{Add the metrics for cross validation. Use the F\textsubscript{tp,fp} from this link and reference properly, also mention why F\textsubscript{average} is not used [REF: https://www.hpl.hp.com/techreports/2009/HPL-2009-359.pdf]}

\textbf{\textit{Time:}}
To compare the time efficiency, we will report the training time for other approaches and fine-tuning time for PTMs, and the prediction time of all the approaches. 
Prediction time is the time the model requires to process the test data and predict labels.
The time will be measured in seconds and will be reported for each dataset.
\textit{A model is considered to be more time efficient than another if the prediction time is less;} as training a model or fine-tuning a PTM is a one-time process and need not be repeated.

%\textcolor{red}{TODO: add explanation that the prediction time is for the whole dataset, similar to the PTM study for sentiment in SE paper. and also that we sum it up for all the folds.}

\textbf{\textit{Time Change in Percentage:}}
\textcolor{black}{We employ all four Prior approaches on all datasets $D_1$--$D_7$ and choose their best as our \textit{baseline} approach. In addition to the Time mentioned above, we will also report the increase and decrease in percentage for the studied PTMs with respect to the considered baseline for the measured Time. The reason of reporting this change in percentage is that time-duration depends on hardware configuration, and percentage change provides a better understanding about the change in time for the readers.}

\section{Approach for Research Questions}
\label{sec:approach_rq}

The following steps will be taken to answer each research question.

\subsection{\textbf{Approach for Research Question 1}}

For RQ1, we identify the accuracy and efficiency of PTMs compared to Prior approaches. For this RQ, all Prior approaches and PTMs will be trained and tested on all of the datasets separately, \textcolor{black}{as explained in section \ref{experimental-setup}. }
We only consider the reviews (not tweets) from \textit{Dataset 2} for RQ1. For each dataset and approach, all the evaluation metrics described in section \ref{sec:evaluation-metrics} will be reported. 
We highlight the best model for each dataset with highest F1\textsuperscript{avg}\textsubscript{micro} and F1\textsuperscript{avg}\textsubscript{macro}. 
The lowest training/fine-tuning and prediction times will also be highlighted for each dataset, along with the time change in percentages. 
We denote the best performing PTM among others as \texttt{PTM-X} to pursue experiments in other research questions. 
If no PTM achieves the best results in terms of both micro- and macro- F1 score, the PTM that achieves the highest F1\textsuperscript{avg}\textsubscript{micro}-score (denoted as \texttt{PTM-XM1}) and the PTM with the highest F1\textsuperscript{avg}\textsubscript{macro}-score (denoted as \texttt{PTM-XM2}) will be selected to be studied in RQ2. 
%\textcolor{orange}{Our initial results on four PTMs and four datasets shows that PTMs can achieve high accuracy.}

\subsection{Approach for Research Question 2}

In RQ2, we are interested to evaluate the performance of PTMs when they are pre-trained on domain-specific corpus rather than non-domain-specific corpora. 
In RQ2, we will use \texttt{PTM-X} (or \texttt{PTM-X1} and \texttt{PTM-X2}) from RQ1 and all the \textcolor{black}{seven} datasets. 
We can either pre-train \texttt{PTM-X} using just the domain-specific app reviews or combine them with the general domain documents to pre-train \texttt{PTM-X} from scratch. 
In the first approach, \texttt{PTM-X} with the same architecture will be trained from scratch on app review sentences.
But pre-training \texttt{PTM-X} on a small number of domain-specific documents has its disadvantages; the model may overfit our dataset and it can result in performance degradation in downstream tasks. 
By pre-training PTMs using both general and domain-specific datasets, we can avoid this problem.
We use the approach of Wada et al. \cite{wada2020medicalBERT}, where we pre-train the \texttt{PTM-X} simultaneously with both general and domain-specific documents. 
Following the literature \cite{wada2020medicalBERT}, we double the frequency of pre-training domain-specific documents during the optimization on the Masked Language Modeling (MLM) task, which is the task the ALBERT and RoBERTa models are trained on.

\begin{figure}[hbt!]
    \centerline{\includegraphics[width=0.4\linewidth]{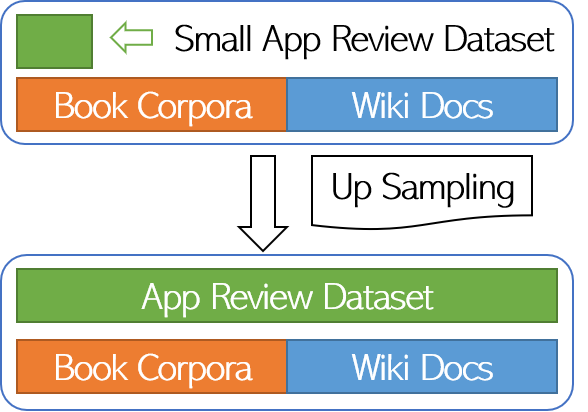}}
    \caption{Creating Up-sampled dataset for PTM pre-training tasks}
    \label{fig:up-sampling}
\end{figure}

{To accomplish this, we must increase the frequency of pre-training for MLM over app reviews. %and use the negative instances of NSP, in which a sentence pair is constructed by pairing two random sentences from different documents \textcolor{red}{one from the general corpora and one from the app reviews dataset}. 
The objective can be achieved by improving the vocabulary representation of app reviews. Therefore, {we use simultaneous pre-training after the up-sampling introduced in} \cite{wada2020medicalBERT}. We create pre-training instances from a set of corpora with different sizes to pre-train a PTM, as described in Figure \ref{fig:up-sampling}. {Upsampling refers to the copying and pasting the app reviews portion of the corpus until the predetermined ratio of app reviews and general pre-training data is achieved; also, we have uniformly mixed the book documents, wiki documents, and app reviews consecutively to form the pre-training corpus.} With this up-sampling technique, more instances from the small app review corpus are used for MLM. }
%and it produces an enhanced number of distinct combinations for NSP.}

We will pre-train three different \texttt{PTM-X}s, where we will respectively integrate 2.8 million, 5.6 million, and 10 million app reviews that we have collected from Google Play, with the general domain documents from Wiki-texts\footnote{The Wiki-texts dataset is open sourced at https://dumps.wikimedia.org/.}. 
Our collected dataset has %\textcolor{blue}{408.6 million app reviews} 
app reviews for more than 2000 apps from different categories. The dataset includes app\_name, app\_category, review, rating, reply\_text, and date.
This will give us three versions of a domain-specific pre-trained model which we refer to as \textit{Customized PTM-X} (\texttt{CPTM-X}).
We denote the three sizes of the \texttt{CPTM-X} as \texttt{CPTM-X\textsubscript{base}}, \texttt{CPTM-X\textsubscript{medium}}, and \texttt{CPTM-X\textsubscript{large}}, respective to the number of the app review sentences integrated for pre-training. 
Finally, we will fine-tune all the \texttt{CPTM-X}s on the \textcolor{black}{seven} datasets according to the steps and evaluation metrics explained in sections \ref{experimental-setup} and \ref{sec:evaluation-metrics}. The best \texttt{CPTM-X} will be used in RQ3. 

\subsection{Approach for Research Question 3}

\textcolor{black}{We have used both the readily available PTMs and our CPTM-X in different settings for evaluating RQ3.}
We are considering four different settings to determine the PTMs' capacity regarding app review analysis. 

\subsubsection{Binary vs. Multi-Class Setting}

\textcolor{black}{Previous studies have shown that the pre-trained Transformer based models deviate from their original performance when the classification task involves multiple classes instead of binary classes \cite{adhikari2019docbert, chang2019x}. 
\textcolor{black}{Studies on app review classification also show that binary classification yields better results \cite{Jha2019MiningNR, maalejea}}.
We are, therefore, interested in evaluating the performance of the PTMs in both binary and multi-class settings, especially, investigating the performance of \texttt{CPTM-X}.}
We will investigate the accuracy and time efficiency of PTMs for binary classification compared to the original multi-class settings. 
We convert all the {seven} datasets into binary-class datasets by randomly choosing one class (i.e. label) for each dataset and switching all the other labels in that dataset into "Others". 
Then, we will use these modified datasets to regenerate the PTM and CPTM results for the binary classification task on each dataset. 
The four PTMs and \texttt{CPTM-X} will be used to record the evaluation metrics. All other steps will remain similar to RQ1.

\subsubsection{Zero-Shot Classification Setting}

The Zero-Shot Learning (ZSL) that we will investigate in our study is the recent approach in which we evaluate the models on fully-unseen labels. 
%In this task, the model is trained on one set of labels and then evaluated on a different set of labels that the classifier has never seen before.
In this setting, we assume that the system is never trained on any labeled data for the task we are interested in, which is different from some ZSL settings in which labels are partially seen by the model. The setting we choose here is more realistic and can provide insights about the potential benefits of PTMs in practice, where the developers might want to use the same model to derive new unseen aspects of the dataset. 
Using PTMs for this kind of zero-shot classification requires different training. 

%In a nutshell, pre-trained natural language inference models (entailment models) can be used to build zero-shot classifiers \cite{zero}. 
For ZSL, we will explore the method proposed by Yin et al. \cite{zero} in which classification is considered as a Natural Language Inference (NLI) task. 
This approach determines the compatibility of two distinct sequences by embedding both sequences and labels into the same space. 
In NLI, a pair of sentences are considered: ``premise" and ``hypothesis"; and the task is to predict whether the ``hypothesis" is an entailment of ``premise" or not (contradiction). 
We follow the steps in \cite{zero} to prepare the datasets and models and set up our study for ZSL. We will use the reviews as ``premise" and candidate labels as ``hypothesis". 
For Example, for the app review ``Please add a back button in the armory page." and its label ``Feature Request", we use the review as a \textit{premise} and the label as \textit{hypothesis}.
So, the NLI model can predict whether the hypothesis is an entailment of the premise or not. This prediction will be compared to the actual labels for the reviews to calculate the evaluation metrics. 
If the NLI model correctly predicts the \textit{hypothesis} as an {entailment} of the \textit{premise}, it is recorded as a {True Positive}. On the other hand, if the NLI model predicts the \textit{hypothesis} as {contradiction} or {neutral} to the \textit{premise}, it is recorded as a {False Negative}.
\textcolor{black}{If the NLI model predicts the \textit{hypothesis} as contradiction or neutral, where the \textit{hypothesis} is actually an entailment, we refer to this as False Positive (FP). On the other hand, if the NLI model predicts the \textit{hypothesis} as contradiction or neutral, where the \textit{hypothesis} is actually contradiction or neutral, it is regarded as True Negative (TN).}

We fine-tune \texttt{CPTM-X} on NLI dataset\footnote{https://cims.nyu.edu/~sbowman/multinli/} and test it on the \textcolor{black}{seven} datasets to determine how well PTMs can classify issues from app-reviews without being trained on them. 
\textcolor{black}{Note that Multi-NLI data does not contain app reviews. }
In addition, we use {\fontfamily{qcr}\selectfont RoBERTa-large-nli}\footnote{https://huggingface.co/joeddav/xlm-roberta-large-xnli}, which is a readily available fine-tuned model for zero-shot classification \footnote{We only use the readily available PTMs for this study and avoid fine-tuning other PTMs on Multi-NLI data in this setting, due to high computational expenses.}. 
\textcolor{black}{This will provide insights about using a custom PTM trained on app reviews and the PTM trained on general purpose corpora in a zero-shot setting.}

%\textcolor{red}{Add this change to the MSR RR differences: In the ZSL, we mentioned PTM-X will be fine-tuned on MultiNLI dataset, but we actually meant CPTM-X. REmoved the N-shot classification from the comparison part of the ZSL. Also, that the results of this ZSL is compared against the RQ1 data. We can also compare the resutls with ARMiner.}

%\textbf{\textit{BART:}} The Bart model uses a standard seq2seq architecture with a bidirectional encoder and a left-to-right decoder.  The pretraining task involves randomly shuffling the order of the original sentences and a novel in-filling scheme, where spans of text are replaced with a single mask token. BART is particularly effective when fine-tuned for text generation but also works well for comprehension tasks. It matches the performance of RoBERTa with comparable training resources on GLUE and SQuAD, achieves new state-of-the-art results on a range of abstractive dialogue, question answering, and summarization tasks.

%\textcolor{brown}{Explain why we dont need app issue classification for these settings (the results are in RQ1).}

\subsubsection{Multi-Task Setting}
\textcolor{black}{Research has confirmed that when a new classification task (e.g. sentiment classification) is introduced, a new model should be trained and the models for another classification task (e.g. app issue classifiers) cannot be applied for this new task {\cite{stanik}}. 
In addition, the same model or classification technique can have different performances on various analysis tasks. For example, Hemmatian and Sohrabi found Naive Bayes (a probabilistic classifier) to be performing better than Decision Tree (a non-probabilistic classifier) for sentiment classification tasks \cite{hemmatian2019survey}; on the other hand, Maalej et al. reported that using Bag-of-Word technique and Decision Tree outperformed Naive Bayes for app issue classification task \cite{maalejea}.
Therefore, we are interested in evaluating the performance of PTMs in both app issue classification and sentiment analysis task settings.
In sentiment analysis, the task is to classify the sentiment of a given review into positive, neutral, or negative polarities \cite{how-far-ptm}, which is different from the app issue classification.}
In this setting, we will evaluate the \texttt{PTMs} and \texttt{CPTM-X} on the new task of sentiment classification for app reviews. 
\textcolor{black}{We fine-tune these models on the sentiment classification dataset and test them for sentiment classification. 
%\textcolor{red}{As we already have these models from RQ1 and RQ2, we test them on sentiment data for this setting.} 
This set up will help evaluating whether the models trained for app issue classification can be applied for another task, thus providing insights about the multi-task ability of the PTMs.}
\textcolor{black}{For the sentiment classification, we will use another dataset\footnote{https://www.kaggle.com/lava18/google-play-store-apps} that consists of 37,185 app reviews with three sentiment polarity labels: \textit{positive} (13,758), \textit{neutral} (9,993), and \textit{negative} (13,434)}. 

We have gone beyond sentiment classification and implemented the PTMs for category classification as well. The same Kaggle datasets also contains category information for 20,955 app review. The number of user reviews in this categories are: \textit{Family} (18.9\%), \textit{Games} (50.3\%), \textit{Tools} (12.5\%), \textit{Medical}  (10.7\%), and \textit{Business} (7.6\%). 
As it is an imbalanced dataset, we use stratified K-fold cross validation for maintaining uniformity with other experiments.
The experimental settings and evaluation metrics we will use are similar to RQ1.
The results will determine how these models perform for a different classification task in the same domain (app reviews).

%\textcolor{red}{We use SentiCR [REF] TODO: AND EXPLAIN WHAT THIS IS as the baseline and also evaluate the Prior approaches for when they are trained and tested on the sentiment classification to assess their usage in multitask setting.}

\subsubsection{Multiple Resources}
The distribution of the data from multiple resources varies and models built for App Stores are not suitable to classify user feedbacks from another resource such as Twitter \cite{SUR-citer7}. In this setting, we will study the accuracy and time efficiency of the PTMs for the classification of data collected from multiple resources. 
For this experiment, \textit{Dataset 2} will be used as it contains App Store reviews and Twitter data with the same labels. 
First, we will fine-tune the \texttt{PTM-X} and \texttt{CPTM-X} on App Store reviews of \textit{Dataset 2} and report the evaluation metrics (training steps are similar to RQ1) when tested on Twitter data.
{In addition, we will fine-tune the \texttt{PTM-X} and \texttt{CPTM-X} on Twitter part of \textit{Dataset 2} and report the results when tested on App Store reviews. 
This way, we will gain insights about the PTMs' capacity to classify app reviews collected from different resources on which the PTMs were not fine-tuned on.
\textcolor{black}{The evaluation metrics used here are explained in section \ref{sec:evaluation-metrics}.}
}
\textcolor{black}{To have a comparison, we also test the Prior approaches for this setting. From the {Prior} approaches, {AR-Miner} does not require training. For the other approaches, we train them on the App Store reviews of \textit{Dataset 2} and test them on Tweets in \textit{Dataset 2}.}
%\textcolor{red}{We only apply this setting for these approaches, as we already know they are developed for classification of app reviews not for Twitter. So, we do not train them on Twitter data and test them on app reviews.}}

\section{Results} \label{sec:results}
In this section, we present the results of our experiments for each of the research questions.
Note that we have computed all the scores (micro- and macro- scores of P, R, F1) and the training and prediction times for all of the models for each dataset. But due to the large number of experiments, we bring all tables in the appendix. In the following, we only discuss the the F1- scores in addition to the prediction time. The full results are shown in \textcolor{black}{Tables \ref{table:RQ1_prior} and \ref{table:RQ1_ptm} for RQ1, Tables \ref{table:RQ2_cptm_albert} and \ref{table:RQ2_cptm_roberta} for RQ2, Tables \ref{table:RQ3_binary_a} and \ref{table:RQ3_binary_b} for RQ3.1, Table \ref{table:RQ3_zero_shot} for RQ3.2, Table \ref{table:RQ3_multi_task} for RQ3.3, and Table \ref{table:RQ3_multi_resource} for RQ3.4 in the appendices}.

\subsection{RQ1: Accuracy and Efficiency of PTMs Compared to Prior approaches}

\begin{figure}[hbt!]
    \centerline{\includegraphics[width=1.0\linewidth]{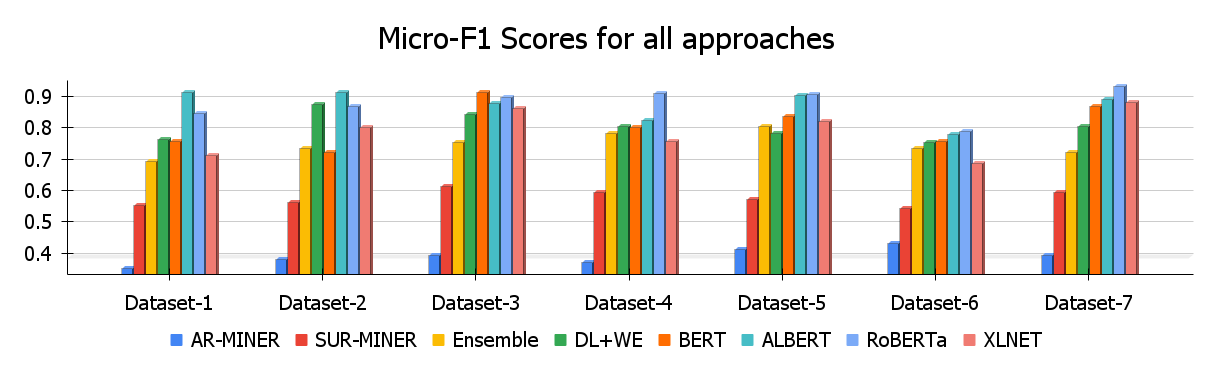}}
    \caption{PTM and Prior Performance (micro-F1 score)}
    \label{fig:ptm_prior_performance}
\end{figure}

Fig \ref{fig:ptm_prior_performance} shows the \textcolor{black}{micro-F1 scores} for Prior approaches and PTMs, where the performance of all models are shown for each dataset.
Among the Prior approaches, \textit{DL+WE} is the best performing model, followed by \textit{Ensemble} approach. Only for $D_5$, which has the lowest number of app-reviews, the \textit{Ensemble} method has better scores than \textit{DL+WE}.
We conjecture that this low number of reviews favored the \textit{Ensemble} model to enhance the prediction performance of participating classifiers.
For the cumulative dataset (\textit{Dataset 7}), \textit{DL+WE} improves the results of \textit{Ensemble Method}, \textit{SUR-Miner} and \textit{AR-Miner} by $\sim$8\%, $\sim$20\%, and $\sim$40\%, respectively. So, we have selected \textit{DL+WE} as our baseline from Prior approaches.

Among the PTMs, on most datasets \textit{ALBERT} and \textit{RoBERTa} achieve the highest F1-micro and F1-macro scores. 
The F1-micro results of these PTMs are higher than the best Prior Approach and other PTMs for all datasets, except that BERT outperforms other models for $D_3$.
The \textit{Ensemble} and \textit{DL+WE} approaches can match the performance of some worse performing PTMs in the respective Dataset, such as \textit{DL+WE} outperformed \textit{BERT} on \textit{Dataset 2}.
The other Prior approaches (i.e., AR-Miner and SUR-Miner) have significantly lower scores than the PTMs for all Datasets.
The performance of \textit{XLNET} varies compared to Prior approaches. For some datasets such as $D_1$, $D_4$, and $D_5$, it has comparable or lower micro-F1 compared to \textit{Ensemble} and \textit{DL+WE}; but for $D_7$ it has higher scores.
\textit{RoBERTa} performs best on all metrics for \textit{Datasets 4 and 6}, and yields best micro-average scores for $D_5$. These three datasets have the least number of app reviews. 
On the other hand, \textit{ALBERT} performs best on  Datasets $D_1$ and $D_2$, the two datasets with the highest number of app reviews. For the cumulative dataset (Dataset 7), \textit{RoBERTa} outperforms the other PTMs on micro-scores, \textit{XLNET} has the highest macro-scores, and the other PTMs have relatively close results. 

For $D_3$, the established order of PTM's performance persists (\textit{RoBERTa} \textgreater \textit{ALBERT} \textgreater \textit{XLNet}), except for the performance yielded by \textit{BERT}. \textit{BERT} outperformed the other three PTMs by a small margin for this dataset ($\sim$1.5\% to $\sim$4.5\%). 
\textcolor{black}{$D_3$ has 4,000 records with six labels, for which the `Other' class has more than 2,000 records. In other words, the models learn about 5 classes from a small portion of reviews. When we further analyzed the confusion matrices of ALBERT and RoBERTa for $D_3$, we find that the main reason for dropping their performance in this dataset is that they predict the records of the 'Other' class wrong, but their prediction for the other five classes are mostly correct. {So, the False Negative predictions are higher and this is also confirmed by having higher precision than recall for these models in $D_3$.} This is in contrast for $D_1$. \textit{Dataset 1} has a large number of records (15,290) labeled as 'Other' out of 34,000 records. The rest of this dataset is labeled by 4 other classes. Therefore, in $D_1$, the number of available reviews per label is much higher than the number of reviews for each label in $D_3$, and the two PTMs perform better as they learn the features for each class. That being said, still, ALBERT and RoBERTa have F1-micro scores of over 85\% and F1-macro of over 83\% for $D_1$ and $D_3$. The score of most models for $D_6$ is their lowest. This is also due to the specific characteristics of this dataset, where it has the lowest number of records per label.}

As both \textit{RoBERTa} and \textit{ALBERT} have the highest or second highest scores in maximum number of datasets, we choose these two PTM models as the best performing ones to pursue with the next research questions. Here, these two PTMs will be our PTM-X models %circumvent an internal threat to validity. 
and are chosen for training from scratch to develop \texttt{CPTM-X (Custom PTMs)}. We refer to these Custom PTMs as C-ALBERT and C-RoBERTa.

\begin{figure}[hbt!]
    \centerline{\includegraphics[width=1.0\linewidth]{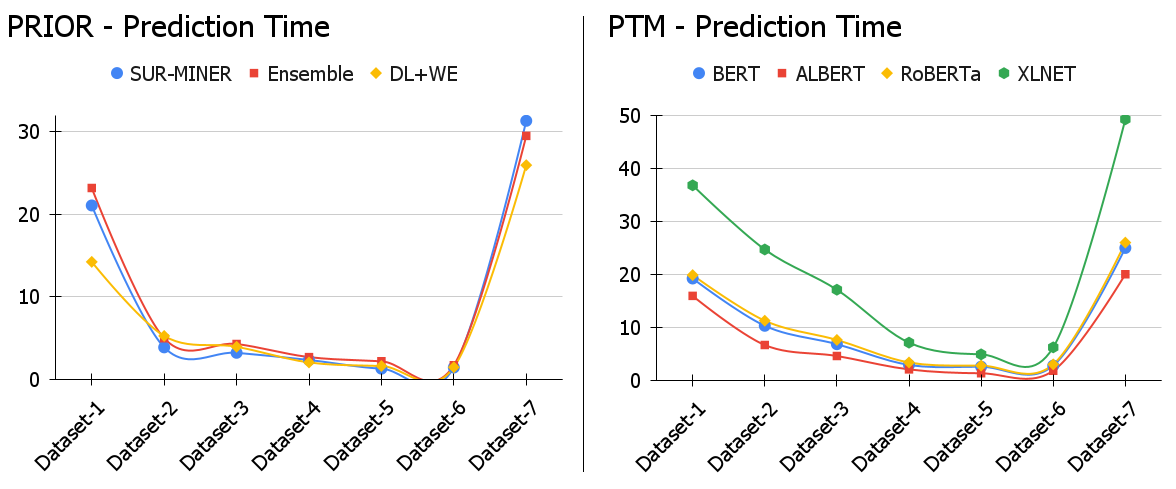}}
    \caption{PTM and Prior Prediction Time (seconds)}
    \label{fig:ptm_prior_pred_time}
\end{figure}

Fig \ref{fig:ptm_prior_pred_time} illustrates the prediction time for Prior and PTM approaches {in seconds}. \textit{AR-Miner} uses topic modeling to predict the labels/classes; so, the prediction time is not reported for \textit{AR-Miner}. From Fig \ref{fig:ptm_prior_pred_time}, we can see that both Prior and PTM approaches follow a similar trend, and their prediction times are related to the size of the dataset, {i.e., the larger the size of the dataset, the longer the prediction time is}.

Among the Prior approaches, our previously selected baseline \textit{DL+WE} yielded the least prediction time on 3 datasets ($D_1$, $D_4$, and $D_7$), whereas \textit{SUR-Miner} method's prediction time is the least for rest of the datasets ($D_2$, $D_3$, $D_5$, and $D_6$). 
Among the PTMs, \textit{ALBERT} consistently yields the least prediction time for all the datasets. On the other hand, the prediction time for \textit{XLNet} is the highest for all datasets.

\textcolor{black}{In the last column of Table \ref{table:RQ1_ptm} (\textit{Time-Diff (\%)}), we have included the percentage change of the prediction times of PTMs with respect to the prediction time of the Prior approach that has the lowest time.
For example, for $D_1$, the Time-Diff (\%) for PTMs are calculated based on the prediction time of \textit{DL+WE}; and for $D_2$ it is calculated based on the prediction time of \textit{SUR-Miner}.
For all datasets except $D_7$, the prediction time of PTMs compared to the best Prior Approach increases by 0.49\% to $\sim$500\%. For $D_7$, \textit{BERT} and \textit{ALBERT} predicted the classes $\sim3\%$ and $\sim22\%$, respectively, faster than the fastest Prior approach (\textit{DL+WE}). For $D_7$, RoBERTa increases the time by 0.15\% and XLNET increases the time by $\sim$89\%.}

% \textcolor{red}{TODO: Add a text box summarizing the results of each RQ. We should clearly mention which of them models are better, in terms of performance and time.}

\vspace{3mm}
\begin{tcolorbox}[colback=black!5!white,colframe=white!50!black,title=Findings of RQ1]
  PTMs are generally superior to Prior approaches when predicting classes with higher scores across all datasets. The two candidates for the best PTMs we choose are ALBERT and RoBERTa, as they invariably perform well across all dataset sizes. Based on the time taken for prediction, DL+WE took the shortest amount of time among all Prior and PTM approaches, and ALBERT has taken the shortest amount of time among the PTM approaches.
\end{tcolorbox}

\subsection{RQ2: Domain-Specific PTMs vs. General PTMs}

\begin{figure}[htbp]
    \centerline{\includegraphics[width=1.0\linewidth]{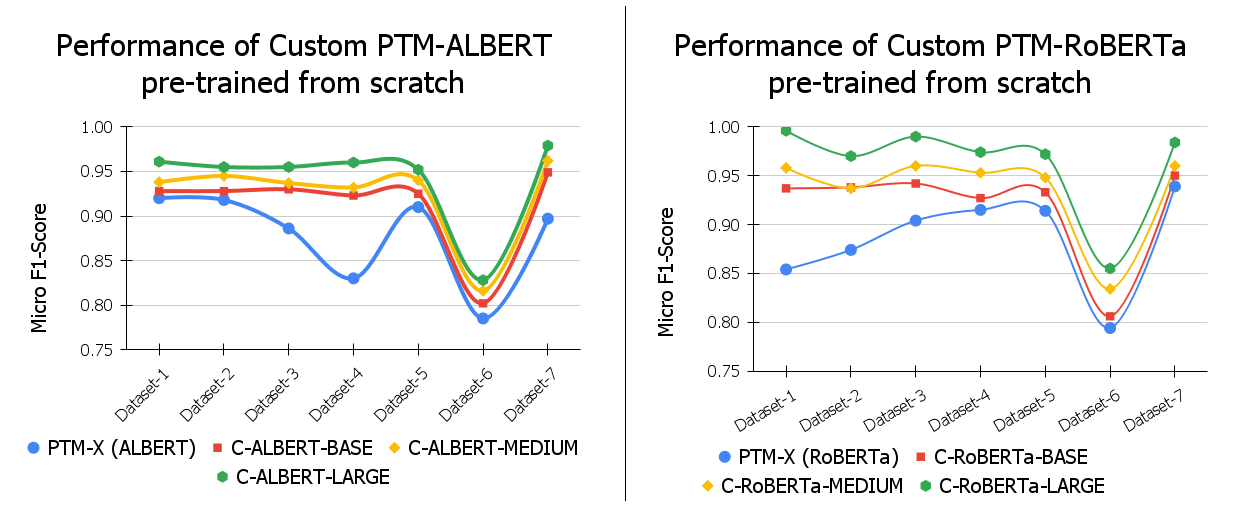}}
    \caption{Performance of Custom PTMs (micro F1 score)}
    \label{fig:cptm_performance}
\end{figure}

Fig \ref{fig:cptm_performance} shows the performance of \textit{C-ALBERT} and \textit{C-RoBERTa} after they are pre-trained on domain-specific app-reviews along with general domain corpora. In \texttt{CPTM-X-Base}, we have incorporated 2.8 million app reviews during pre-training. For \texttt{CPTM-X-Medium} and \texttt{CPTM-X-Large}, we have increased the number of app reviews during pre-training to 5.6 million and 10 million, respectively.
From the result, we can see that all the Custom PTMs performed better in all datasets than their corresponding out-of-the-box PTMs. We also find that increasing the number of app reviews in pre-training corpora helps boost the performance of Custom PTMs.

\textit{ALBERT} has lower F1-micro scores for datasets $D_3$ and $D_4$. After pre-training with app reviews, we notice that all \textit{C-ALBERT} models (i.e., C-ALBERT-BASE, C-ALBERT-MEDIUM, and C-ALBERT-LARGE) have significantly better predictions for both of these datasets ($\sim5\%$ to $\sim13\%$ improved performance). Although this is not the case for $D_6$, all \textit{C-ALBERT} models still improve their performances. We relate this to the low number of reviews per class in $D_6$. The Custom PTMs produce better predictions than their readily available counterpart. 
\textit{C-RoBERTa} models also have better performance, yielding up to 15.2\% increase in performance over \textit{RoBERTa}.
As seen in Fig \ref{fig:cptm_performance}, \textit{C-RoBERTa} improves RoBERTa's F1-micro scores for datasets $D_1$--$D_4$, where \textit{RoBERTa} has scores of below 95.
There is none to little fluctuation in prediction time for all the Custom PTMs.

\vspace{3mm}
\begin{tcolorbox}[colback=black!5!white,colframe=white!50!black,title=Findings of RQ2]
    Incorporating domain-specific data (i.e., app reviews) during pre-training of the PTMs improves their performance for app review classification, and slightly improves the prediction time of CPTMs compared to PTMs. 
    The performance of the models benefits from incorporating more domain-specific data in the pre-training.
\end{tcolorbox}

\subsection{RQ3: Experimenting PTMs in Different Settings}

\subsubsection{RQ3-1: Binary vs. Multi-class Setting}
    We report the binary classification results in Tables \ref{table:RQ3_binary_a} and \ref{table:RQ3_binary_b} for RQ3-1. The multi-class classification results of the PTMs and CPTMs are previously reported in Table \ref{table:RQ1_prior} to Table \ref{table:RQ2_cptm_roberta}. The micro- and macro- scores were calculated for accommodating the existence of multi-class in our datasets, which is no longer the case after we converted each multi-class datasets into binary class datasets.
    
    The randomly selected classes for \textit{$D_1$} to \textit{$D_7$} are \textit{Bug-Report, Inquiry, Reliability, Feature Requests, App Problem, Bug-Report, and Aspect Evaluation}, respectively. All the other classes in these datasets are labeled as \textit{Other} or \textit{Irrelevant}.
    
    From the results in Table \ref{table:RQ3_binary_a} and Table \ref{table:RQ3_binary_b} we observe that both \textit{Prior} and \textit{PTM} approaches have performed better in binary classification compared to multi-class classifications. The improvement for the \textit{Prior} approaches ranges from 0 (for \textit{AR-Miner}) to $\sim$3.2\%. On the other hand, the improvement for the \textit{PTM} approaches ranges from 1 to $\sim$7.7\%. All the PTM approaches yielded better performance for binary classification than they did for multi-class classifications and we observe that readily-available PTMs' performance boost is similar to the performance boost produced by the Custom PTMs.

\subsubsection{RQ3-2: Zero-Shot Setting}
    We report \textit{Precision}, \textit{Recall}, and \textit{F1 score} for zero-shot classification results in Table \ref{table:RQ3_zero_shot}. %The micro- and macro- scores were calculated for accommodating the existence of imbalanced multiple classes in our datasets, which may have imbalanced contribution to the fine-tuning training set and testing set. 
    As in this setting, we are not fine-tuning our model on a portion of our considered Datasets \textit{$D_1$} to \textit{$D_7$} (rather, we are fine-tuning our PTMs on a separate NLI dataset), we no longer require cross validation over multiple folds.
    
    In this setting, we anticipate lower performance because our PTM models do not have any exposure to any instances of the datasets to fine-tune on. 
    {In this experiment, we use AR-Miner as the baseline to compare the results of this setting for PTMs and CPTMs. AR-Miner is chosen as it is a topic modeling based approach that does not require seeing the label beforehand. The other models from Prior approaches are not considered here as all of them are supervised machine learning or deep learning techniques that require training with the labels beforehand.}
    The \textit{C-RoBERTa-LARGE-NLI} and \textit{C-ALBERT-LARGE-NLI} are the CPTMs that are pre-trained on 10 million app reviews (See RQ2).
    Overall, in \textit{$D_7$}, \textit{RoBERTa-LARGE-MNLI} yielded best performance (71\%). 
    We observe that RoBERTa-based models achieve higher scores compared to ALBERT-based PTMs. 
    Results of AR-Miner vary between 36\% and 44\% F1 score for the datasets,  and AR-Miner has closer results to \textit{ALBERT-NLI}. 
    For \textit{$D_5$}, \textit{ALBERT-NLI} yielded a $\sim$37\% F1 score, which is $\sim$5\% worse than the performance of AR-Miner. For Datasets \textit{$D_1$} to \textit{$D_7$}, the PTMs yield on an average 45.875\%, 44.425\%, 43.125\%, 43.225\%, 46.45\%, 38.8\%, 42.725\% less F1 score than their counterparts in RQ1. All PTM approaches outperformed AR-Miner, the only prior approach capable of zero-shot classification, except in \textit{$D_5$} for \textit{ALBERT-NLI}.
    The Custom PTMs in all cases improve the results by approximately 10 F1 scores in all datasets.
    Similar to previous RQs, the ALBERT-based model has a lower prediction time.

\subsubsection{RQ3-3: Multi-task Setting}
    In Tables \ref{table:RQ3_multi_task} and \ref{table:RQ3_multi_task_sentiment}, we report the micro and macro \textit{Precision}, \textit{Recall}, and \textit{F1 score} and the training and prediction times for two classification tasks in the app review domain: category classification and sentiment classification. %We have discarded macro-average \textit{Precision}, \textit{Recall}, and \textit{F1-Score} for brevity as we have only used micro-average score to compare PTMs' performances. 
    
    Among all the PTMs, C-ALBERT-LARGE yielded the best result for category classification tasks with micro-average F1 scores of $\sim$91\% and C-RoBERTa-LARGE yielded the best result for sentiment classification tasks with micro-average F1 scores of $\sim$93\%, respectively. 
    
    BERT, ALBERT, RoBERTa, and XLNET performed 2.4\%, 1.8\%, 7.5\%, and 4.3\% worse for category classification, respectively; they performed 1.1\%, 0.1\%, 5.5\%, and 0.9\% worse for Sentiment classification task, respectively with respect to their averaged micro-F1 scores over 7 datasets, in Table \ref{table:RQ1_ptm}. C-ALBERT-BASE, MEDIUM, and LARGE performed 4\%, 2.7\%, and 2.4\% worse for category classification, respectively, compared to their averaged micro-F1 scores over 7 datasets, in Table \ref{table:RQ1_ptm}; they performed 2.8\%, 3.4\%, and 2.3\% worse for the sentiment classification task, respectively with respect to their F1 scores averaged over 7 datasets, in Table \ref{table:RQ2_cptm_albert}. On the other hand, C-RoBERTa-BASE, MEDIUM, and LARGE performed 4.1\%, 5.9\%, and 6.2\% worse for category classification, respectively; they performed 1.5\%, 2.3\%, and 3.3\% worse for the sentiment classification task, respectively with respect to their F1 scores averaged over 7 datasets, in Table \ref{table:RQ2_cptm_roberta}. %We can observe that all these performance reductions are marginal and better than the performances of Prior approaches.
    For both classification tasks, we observe that the custom PTMs achieve higher scores compared to the PTMs, showing that although the task is different (e.g., sentiment analysis vs. app review classification), they still benefit from pre-training the models on the same domain (app reviews). 

\subsubsection{RQ3-4: Multi-resource Setting}
    Dataset \textit{$D_2$} contains labeled app reviews from two different sources: App Store and Twitter. In the previous results (Table \ref{table:RQ1_prior} to \ref{table:RQ3_multi_task}), we have not incorporated app reviews collected from Twitter. Here, we use the reviews from both App Store and Twitter from \textit{$D_2$}.
    
    \textbf{Training on App Store App-reviews and Testing on Twitter.} In the first part of this setting, we fine-tuned all the available approaches (except for AR-Miner) on app reviews from App Store and tested them on app reviews collected from a different resource, i.e., Twitter. The results are available in Table \ref{table:RQ3_multi_resource}. Being a topic modeling approach, \textit{AR-Miner} does not require any supervised training.
    
    C-RoBERTa-LARGE achieves the best score among all the considered approaches for this setting, in terms of F1 score (0.86).
    From the previous approaches, DL+WE approach scored higher than the readily available PTMs, and all Custom PTMs, except C-RoBERTa-LARGE. It also beats the C-ALBERT-LARGE by 1\% F1 score. The Ensemble method surpassed only readily available PTMs and closely matched the performance of C-ALBERT-MEDIUM.
    AR-Miner and SUR-Miner have the lowest scores among the Priors, and, we do not observe a big reduction in their performance with respect to single-resource setting, in terms of F1 scores. These approaches have scored  ($\sim$5\%) and ($\sim$2\%) less in multi-resource setting than their single-source counterparts.
    
    Compared to the setting where PTMs are evaluated only on app reviews from App Store, we observe $\sim$13\% to $\sim$31\% performance reduction for micro-F1 scores. This performance reduction is less for Custom-PTMs. % showed promise in this setting by producing micro-F1 scores that suffer comparatively less performance reduction from $\sim$15 to $\sim$25\%. 
    Among the Custom ALBERT models, the \textit{C-ALBERT-LARGE} model produced the best micro-F1 score of 0.80 ($\sim$25\% reduction) and \textit{C-ALBERT-BASE} model produced the least micro-F1 score of 0.71 ($\sim$21\% reduction). The C-RoBERTa group has less reduction compared to C-ALBERT models, where the \textit{C-RoBERTa-LARGE} produced the best micro-F1 score of 0.86 ($\sim$15\% reduction) and \textit{C-RoBERTa-BASE} model produced the least micro-F1 score of 0.77 ($\sim$15\% reduction) among all C-RoBERTa models.
    
    \textbf{Training on Twitter and Testing on App Store App-reviews.} \textcolor{black}{In the second part of this setting, we trained/fine-tuned all the available approaches (except for AR-Miner) on Twitter and tested them on app reviews collected from a different resource, App Store. The results are available in Table \ref{table:RQ3_multi_resource_2}.}
    
    \textcolor{black}{C-RoBERTa-LARGE achieves the best score among all the considered approaches for this setting, with an F1 score of 0.92.
    From the previous approaches, the DL+WE approach scored higher than the readily available PTMs in this setting. Here, it could not surpass the C-ALBERT-LARGE but beat the C-ALBERT-MEDIUM by 1\% F1 score. The Ensemble method outperformed only readily available PTMs and  C-ALBERT-BASE from the Custom PTMs.
    AR-Miner and SUR-Miner have the lowest scores among the Priors, which have scored  ($\sim$10\%) less and ($\sim$2\%) more in multi-resource setting than their single-source counterparts, respectively. SUR-Miner beat the BERT in this setting by $\sim$4\%.}
    
    \textcolor{black}{Compared to the setting where PTMs are evaluated only on app reviews from App Store, we observe $\sim$9\% to $\sim$30\% performance reduction for micro-F1 scores. This performance reduction is less for Custom-PTMs. % showed promise in this setting by producing micro-F1 scores that suffer comparatively less performance reduction from $\sim$15 to $\sim$25\%. 
    Among the Custom ALBERT models, the \textit{C-ALBERT-LARGE} model produced the best micro-F1 score of 0.87 ($\sim$8\% reduction), and \textit{C-ALBERT-BASE} model produced the least micro-F1 score of 0.74 ($\sim$18\% reduction). The C-RoBERTa group has less reduction compared to C-ALBERT models, where the \textit{C-RoBERTa-LARGE} produced the best micro-F1 score of 0.92 ($\sim$5\% reduction) and \textit{C-RoBERTa-BASE} model produced the least micro-F1 score of 0.78 ($\sim$14\% reduction) among all C-RoBERTa models.}
    The smaller performance reduction for the models in this part, compared to when models are trained on App Store data and tested on Twitter data, could be related to the size of the training data. The dataset $D_2$ has a higher number of labeled \textcolor{black}{tweets ($\sim$ 10K) than the app reviews from App Store ($\sim$ 6.5K). }
    
    The prediction times of all Custom PTMs are lower than Prior approaches in this setting.
    
    %The Custom PTMs perform better than readily available PTMs as the Custom PTMs acquire more syntactic and semantic knowledge about the domain during their pre-training. As app reviews have some unique attributes (e.g., shortness of texts), the domain specific knowledge helps the Custom PTMs to better classify the texts related to the domain they are trained on.

    \vspace{3mm}
\begin{tcolorbox}[colback=black!5!white,colframe=white!50!black,title=Findings of RQ3]
\begin{enumerate}
    \item Both Prior and PTM models yield better performance for binary classification tasks than they do for multi-class setting.
    
    \item PTMs are the best choice for zero-shot classification setting and Custom PTMs, though having less score than the large RoBERTa-MNLI model, improve the results of their non-domain specific models by approximately 10 F1 scores. %and better than readily available topic modeling techniques, such as AR-Miner.
    
    \item For multi-task and multi-resource settings, the readily available PTMs and Custom PTMs showed the same trend as demonstrated in RQ1 and RQ2, that is: \textit{Readily Available PTM} $<$ \textit{Custom-BASE} $<$ \textit{Custom-MEDIUM} $<$ \textit{Custom-LARGE}.
    
    \item RoBERTa-based Custom PTMs have the best scores among all models. They also have lower prediction times in the multi-resource setting compared to Prior approaches.

\end{enumerate}
\end{tcolorbox}

\section{Discussions}
\label{sec:disc}

% \textcolor{red}{What are the implications for researchers, practitioners, and app developers or companies? How the results of this work will help others? We should refer to the motivations of this study to elaborate on these. for example, developers can use the PTMs and apply them directly on their custom classification task, without worries about labeling new data for new classes. }

%In this section, we discuss the lesson learned from from our study and how the findings can be helpful for the researchers, practitioners, and app developers or companies.

\subsection{Implications For Users}

In the following, we discuss our findings and implications for the users\footnote{Here, users can be developers, practitioners, or researchers who want to use the studied models to classify issues related to mobile apps from user feedbacks.}.

\textbf{Prior approaches: \textit{Deep learning with word embedding can be used for app issue classification when one does not want to use PTMs.}} From the Prior approaches, Deep Learning Technique leveraging Word Embedding (DL+WE) generated the best results for predicting the correct classes of app reviews (RQ1), followed closely by the Ensemble Method. DL+WE also took the least amount of time to predict the classes, whereas SUR-Miner took the second least amount of time. However, if no labels are available to train the mentioned models, AR-Miner should be used, which utilizes topic modeling to separate the classes of app reviews.

\textbf{Prior approaches vs. PTMs: \textit{If the prediction time does not matter much but high performance is required, PTMs should be used to classify app issues.}}
From the results of RQ1, we observed that all PTMs outperformed the Prior approaches (except for 1 out of 28 cases) to predict classes accurately. RoBERTa and ALBERT scored the highest F1 scores in most of the datasets among D1 to D7. Compared with the fastest prior approach, ALBERT has the lowest, and XLNET has the highest increase in prediction time. 
Consequently, it is worth applying PTMs for app-review classification if the developer, researcher, or software engineer can spare marginally additional time to achieve higher prediction scores.

\textbf{Prior approaches vs. PTMs: \textit{PTMs can be used for all settings when higher performance is required, except for multi-resource setting, where the deep learning model with word embedding has better scores.}}
For the multi-class classification (RQ1), PTMs perform significantly better than the Prior approaches for most datasets. For Dataset $D_2$, the difference in the scores among the two groups is not much, but when we consider $D_7$, the results of PTMs are almost 10 F1 score higher than the best Prior Approach. The prediction times also vary, but, some PTMs can predict the labels at approximately the same time as the best Prior approaches (Ensemble and DL+WE). 
For binary classification, the best Prior approaches achieve scores of below 90\% and mostly in the low 80\% F1 score. But, the F1 scores of PTMs are mainly above 90\% and the Custom PTMs have even higher scores. 
However, for multi-resource setting (RQ3-4), the Ensemble and DL+WE have better F1 scores than all PTMs, but not higher than CPTMs. This result is interesting, showing that the learned knowledge about the app review domain for issue classification is important in this setting, no matter if the model is trained on App Store data and tested on Twitter data, or vice versa. This is confirmed by the obtained results of the C-RoBERTa-LARGE, which have the highest scores among all models in this setting.

\textbf{Binary vs. Multi-class: \textit{When possible, the classification task should be turned into a binary classification to achieve higher scores.}}
All PTMs and Prior approaches performed better in binary classification than in multiclass classification.
From Tables \ref{table:RQ3_binary_a} and \ref{table:RQ3_binary_b} , we observe that converting or trimming a multi-class dataset to a binary class dataset help all the prior and PTM approaches to yield better performance. Prior approaches are improved by 0 to $\sim$3\%, whereas the gain for the PTMs ranges from 1 to $\sim$7\%. For binary classification, all PTM approaches performed better than they did for multi-class classifications, and the performance boost produced by readily-available PTMs is proportionate to the performance boost produced by Custom PTMs.

\textbf{Zero-shot: \textit{In a zero-shot setting, the Roberta-based Custom PTMs or RoBERTa-LARGE-MNLI can be used to classify app issues.}}
All PTMs can be implemented in a Zero-Shot setting and yield adequate performance.
The only Prior approach that can be used directly in this setting is AR-Miner which is based on topic modeling. 
All the readily available and Custom PTMs are fine-tuned on NLI dataset for Inference classification task. Due to the fact that these PTM models had not been exposed to any instances of the datasets in our study in the zero-shot setting, they are expected to perform worse than their reported performance in Tables \ref{table:RQ1_ptm}, \ref{table:RQ2_cptm_albert}, and \ref{table:RQ2_cptm_roberta}. 
%Except for one case of \textit{$D_5$} where \textit{PTM-ALBERT-NLI} performed worse than AR-MINER, PTMs beat AR-Miner by a large margin in all datasets. Among the PTMs, \textit{RoBERTa-LARGE-NLI} yielded the best F1-scores in all the datasets. 
The Custom-PTMs pre-trained with domain-specific knowledge could not outperform the readily available \textit{RoBERTa-LARGE-MNLI}. This can be related to the dataset used for its training, which is a much larger dataset than the NLI datasets we used to train our PTMs.  
Although the CPTMs cannot beat the \textit{RoBERTa-LARGE-MNLI} model, it is notable that these CPTMs are smaller models and the \textit{RoBERTa-LARGE-MNLI} model is a very large language model; but still, the CPTMs we studied achieve significant improvements compared to their general domain PTMs in this setting.

\textbf{Custom PTMs vs. PTMs: \textit{Custom PTMs are the best models among all that should be used to achieve better prediction scores with lower prediction times.}}
Customized PTMs that are pre-trained with the app review data have the best performance in all settings: binary classification, multi-class classification, multi-task, and multi-resource setting. Even for zero-shot setting, they perform better than all the other models (except for RoBERTa-LARGE-MNLI which is trained on a much larger NLI data). 
The benefit of using Custom PTMs, specifically RoBERTa-based ones, compared to PTMs, is more obvious for the multi-task and multi-resource setting in RQ3-3 and RQ3-4. 
Interestingly, the PTMs do not perform well in the multi-task and multi-resource setting, while CPTMs have the highest scores. The PTMs in these two settings perform lower than the Ensemble and DL+WE models. But, the best CPTMs achieve scores of above 90\% for multi-task and 86\% and 92\% F1 score for multi-resource. 
Other than achieving the best score, CPTMs have the lowest prediction time among all models for multi-class classification, multi-task, and multi-resource settings.
They also achieve the lowest prediction time for some of the datasets in binary classification.
Our results for RQ3-4 confirm that using over-the-shelf PTMs might not be the best option when classifying issues for app reviews in all settings.
PTMs can still have good performance in binary and multi-class classification. But for harder scenarios (i.e., zero-shot, multi-task, and multi-resource), pre-training them with domain-specific data increases their performance and reduces their prediction time. 
Moreover, the more app review data is used in the pre-training of CPTMs, the better their performances are.

\subsection{Implications For Researchers}

\textbf{Training Custom PTMs: \textit{Research for incorporating app review data in the pre-training of the current PTMs is needed.}}
The inclusion of app reviews in the pre-training corpus allows PTMs to classify user feedbacks with higher F1 scores. 
We have pre-trained three models of both ALBERT and RoBERTa with three different sizes of corpora. 
%The BASE, MEDIUM, and LARGE iterations of both ALBERT and RoBERTa incorporate 2.8M, 5.6M, and 10M app reviews to the initial pre-training corpus, respectively.
This experiment found that more domain-specific information during the pre-training helps the newly trained PTMs yield better predictions of app reviews. C-ALBERT and C-RoBERTa models can improve the ALBERT and RoBERTa scores for up to $\sim$13\% and $\sim$14\%, respectively. 
Here, we followed the work of \textcolor{black}{Wada et al. \cite{wada2020medicalBERT}} to incorporate app review data for pre-training of the PTMs. \textcolor{black}{However, there are multiple research on training the PTMs with the domain-specific data \cite{legal-bert, hakala2019biomedical}.} Other approaches can be investigated for domain-specific models that achieve higher scores in the settings of RQ3.
%-BASE, MEDIUM, and LARGE yield $\sim$[0.8-9\%], $\sim$[1-10\%], and $\sim$[4-13\%] increase in performance, respectively; whereas, RoBERTa-BASE, MEDIUM, and LARGE yield $\sim$[1-8\%], $\sim$[3-10\%], and $\sim$[5-14\%] increase in performance, respectively. 
%\textcolor{red}{In all Custom PTMs, there is a lack of significant variance with regard to the prediction time reported for those that do not include app reviews into the pretraining corpus as the prediction time depends solely on the prediction stage rather than the training step.}

\textbf{Multi-class Classification and Zero-shot Settings: \textit{Research needed to build models that perform better in multi-class and zero-shot setting.}}
Our results showed that the studied models perform better in a balanced binary classification setting, rather than an imbalanced multi-class setting. This suggests that more research is needed to develop models that perform better when multiple labels are available and the dataset is imbalanced.
Moreover, researchers can work on a method to increase the performance of the models when classifying with new labels that the model has not seen are required.

\section{Threats to Validity}
\label{sec:threats}

\textbf{Internal Validity.}
    \textcolor{black}{
    A possible internal validity can be related to the obtained results. To mitigate this threat, we use stratified k-fold cross-validation to avoid the bias that might be introduced to the results by the test set. 
    \textcolor{black}{In addition, to mitigate threats to the validity of the results, hyper-parameter values are kept the same for all PTMs and Deep learning models in all the fine-tuning steps. \textcolor{black}{Also, we prevent overfitting by using early stopping and higher dropout rate.} We also keep the values of alpha and beta for LDA as reported in AR-Miner.}  We run all the experiments on a single machine, and we report the machine configuration to enforce the reproducibility of the results. Furthermore, we consider the same metrics to compare the PTMs with Prior approaches. Along with the generated raw duration, we also provide the change of time in percentage as compared to the baseline. Our adopted stricter micro and macro-metrics diminish the ambiguity while providing comparisons among the considered approaches' associated results. }
    
\textbf{Construct Validity.}
    \textcolor{black}{The selection of the Prior approaches and PTMs can pose a validity threat to our study. We identified the four most common approaches as priors by examining the highly practiced methods, tools, and techniques employed by researchers and application developers. The considered approaches include different Machine Learning algorithms, Ensemble methods, and Deep Learning Approaches for app issue classification tasks and are selected after conducting a literature review. %We chose these methods as PRIORs after conducting a comprehensive literature review.
    The PTMs were adopted by following previous study \cite{how-far-ptm} that conducted an empirical study on PTMs performance for sentiment analysis in Software Engineering. Additionally, we pre-trained transformer-based models by incorporating app-review-related datasets with the previously used generic dataset to better understand the potential of domain-specific PTMs (\texttt{CPTM-X}) for app issue classification.
    Another threat to the study can be related to dataset $D_7$. This dataset is a merge of other datasets and has labels borrowed from them. Although we consulted the definitions of the labels from their publishers, there might be a chance that the samples in one group of a dataset are closely related to the samples from another label of another dataset, thus, affecting the results. We mitigate this threat by combining labels from different datasets that have similar names. We also used stratified k-fold cross validation to alleviate potential threats.}

\textbf{External Validity.}
    \textcolor{black}{
    In this study, we empirically study the ability of the PTMs in issue classification of app reviews. The app review classification mainly focuses on extracting useful information in the context of software engineering; which can be used by app developers for requirements engineering, release planning, and other tasks as mentioned in the paper. Our study is therefore limited to app review classification for software engineers and we do not study PTMs for other purposes such as finding issues from business perspectives \cite{tang2019systematicCommerce}, product reviews \cite{zhao2017productReview}, and applications such as intention mining \cite{huang2018intentionMining}. We also do not study the summarization of relevant issues from app reviews. 
    However, we conduct our study on six different datasets. In addition, by merging them, we run experiments on $D_7$ which has multiple classes covering different aspects of apps useful for app developers. 
   % In the future, we plan to curate datasets labeled with security and energy consumption (related to battery issues) issues and conduct the same experiments on these datasets to extend the external validity of the results.
    Another threat in the generalizability of the results lies in different tasks we considered. We mitigate this threat by studying the sentiment classification and category classification of app reviews in the multitask setting. Although PTMs \textit{might} be useful for other tasks, we do not study them in this work. 
    To ensure the generalizability of our study in the specified scope, we have incorporated datasets that include app reviews from diverse app categories and from two platforms i.e., Google Play and Apple App Store. The datasets have various sizes and different labels. In addition, we experiment on issue classification from another platform, i.e. Twitter, which has been shown to require different models than App Stores due to the differences of the platforms, noise data, and user feedback.}

\section{Conclusion} 
\label{sec:conclusion}

We conducted an extensive exploratory study comparing app issue classification tools and pre-trained Transformer-based models in various settings. 
We conducted the experiments on six available datasets and a highly imbalanced dataset, which is a combination of the six datasets. 
Domain-specific PTMs were trained using different sizes of app review data we collected from Google Play and these customized PTMs were also studied here.
Our results confirm that PTMs are achieving higher scores in binary and multi-class classification compared to Prior approaches, but the over-the-shelf PTMs are not always the best models to be used in all scenarios. Instead, CPTMs have the highest scores and are able to perform better than other models in all settings. Moreover, incorporating app specific data in the pre-training of PTMs reduces the prediction time. 
One of the future directions of this research is assessing domain-specific PTMs in other areas of app reviews and exploring ways to increase performance in zero-shot setting.
%We are the first to evaluate the effectiveness of different pre-trained transformer-based models and their variations for the app issue classification task in . %One original dataset, as well as six publicly available datasets consisting of labeled app reviews, were used for our study. 
%Using our experiment results, we identified and reported the best performing fine-tuned Transformer model in terms of the macro-averaged and micro-averaged F1 scores. Overall, we concluded that Transformer-based approaches and their variations outperformed the Prior approaches by a considerable margin for app reviews classification.

%\begin{acknowledgements}
%\end{acknowledgements}



\section{Appendix}
\begin{table}[htbp]
    \caption{RQ1 results for Prior approaches. The best scores for each dataset are shown in bold.}
    \begin{center}
    \renewcommand{\arraystretch}{1.2}
    \resizebox{1.0\linewidth}{!}{%
    \begin{tabular}{l | c | c | c | c | c | c | c | c}
        \hline
	    {} & \multicolumn{3}{c|}{Micro-avg Scores} & \multicolumn{3}{c|}{Macro-avg scores} & \multirow{3}{4em}{Training Time (s)} & \multirow{3}{4em}{Prediction Time (s)} \\
	    \cline{2-7}
        {} & P & R & F1 & P & R & F1 & {} & {} \\
        \cline{1-7}
        \textbf{Dataset-1} & \multicolumn{6}{c|}{} & & \\
        \cline{2-9}								
        AR-MINER & 0.36 & 0.36 & 0.36 & \multicolumn{3}{|c|}{No micro or macro average} & 261.71 & NA \\
        \cline{2-9}
        SUR-MINER & 0.57 & 0.56 & 0.56 & 0.53 & 0.47 & 0.5 & 105.31 & 21.06 \\
        \cline{2-9}
        Ensemble & 0.71 & 0.7 & 0.7 & 0.7 & 0.6 & 0.65 & 115.84 & 23.17 \\
        \cline{2-9}
        DL+WE & \textbf{0.84} & \textbf{0.71} & \textbf{0.77} & \textbf{0.83} & \textbf{0.67} & \textbf{0.74} & 99.62 & \textbf{14.23} \\
        \hline
        
        \textbf{Dataset-2} & \multicolumn{8}{c}{}  \\
        \cline{2-9}	
        AR-MINER & 0.49 & 0.32 & 0.39 & \multicolumn{3}{|c|}{No micro or macro average} & 81.17 & NA \\
        \cline{2-9}
        SUR-MINER & 0.6 & 0.55 & 0.57 & 0.65 & 0.51 & 0.57 & 27.18 & \textbf{3.88} \\
        \cline{2-9}
        Ensemble & 0.8 & 0.69 & 0.74 & 0.7 & 0.66 & 0.68 & 29.9 & 4.98 \\
        \cline{2-9}
        DL+WE & \textbf{0.93} & \textbf{0.84} & \textbf{0.88} & \textbf{0.94} & \textbf{0.84} & \textbf{0.89} & 26.31 & 5.26 \\
        \hline
        
        \textbf{Dataset-3} & \multicolumn{8}{c}{}  \\
        \cline{2-9}	
        AR-MINER & 0.44 & 0.36 & 0.4 & \multicolumn{3}{|c|}{No micro or macro average} & 74.62 & NA \\
        \cline{2-9}
        SUR-MINER & 0.62 & 0.63 & 0.62 & 0.59 & 0.53 & 0.56 & 19.33 & \textbf{3.22} \\
        \cline{2-9}
        Ensemble & 0.77 & 0.75 & 0.76 & 0.69 & 0.71 & 0.7 & 21.46 & 4.29 \\
        \cline{2-9}
        DL+WE & \textbf{0.86} & \textbf{0.84} & \textbf{0.85} & \textbf{0.77} & \textbf{0.72} & \textbf{0.74} & 19.74 & 3.95 \\
        \hline
        
        \textbf{\textcolor{black}{Dataset-4}} & \multicolumn{8}{c}{}  \\
        \cline{2-9}	
        AR-MINER & 0.45 & 0.33 & 0.38 & \multicolumn{3}{|c|}{No micro or macro average} & 71.96 & NA \\
        \cline{2-9}
        SUR-MINER & 0.63 & 0.58 & 0.6 & 0.55 & 0.53 & 0.54 & 11.74 & 2.35 \\
        \cline{2-9}
        Ensemble & 0.84 & 0.74 & 0.79 & 0.82 & 0.7 & 0.76 & 13.5 & 2.7 \\
        \cline{2-9}
        DL+WE & \textbf{0.87} & \textbf{0.76} & \textbf{0.81} & \textbf{0.85} & \textbf{0.73} & \textbf{0.79} & 12.29 & \textbf{2.05} \\
        \hline
        
        \textbf{Dataset-5} & \multicolumn{8}{c}{}  \\
        \cline{2-9}	
        AR-MINER & 0.53 & 0.35 & 0.42 & \multicolumn{3}{|c|}{No micro or macro average} & 57.02 &  NA \\
        \cline{2-9}
        SUR-MINER & 0.62 & 0.54 & 0.58 & 0.57 & 0.56 & 0.56 & 9.11 & \textbf{1.3} \\
        \cline{2-9}
        Ensemble & \textbf{0.86} & \textbf{0.76} & \textbf{0.81} & \textbf{0.84} & \textbf{0.81} & \textbf{0.82} & 10.93 & 2.19 \\
        \cline{2-9}
        DL+WE & 0.84 & 0.75 & 0.79 & 0.83 & \textbf{0.81} & \textbf{0.82} & 9.73 & 1.62 \\
        \hline
        
        \textbf{Dataset-6} & \multicolumn{8}{c}{}  \\
        \cline{2-9}	
        AR-MINER & 0.47 & 0.41 & 0.44 & \multicolumn{3}{|c|}{No micro or macro average} & 71.91 & NA \\
        \cline{2-9}
        SUR-MINER & 0.58 & 0.52 & 0.55 & 0.6 & 0.57 & 0.58 & 10.07 & \textbf{1.44} \\
        \cline{2-9}
        Ensemble & 0.75 & 0.73 & 0.74 & 0.85 & 0.81 & 0.83 & 11.98 & 1.71 \\
        \cline{2-9}
        DL+WE & \textbf{0.77} & \textbf{0.76} & \textbf{0.76} & \textbf{0.86} & \textbf{0.82} & \textbf{0.84} & 10.3 & 1.47 \\
        \hline
        
        \textbf{\textcolor{black}{Dataset-7}} & \multicolumn{8}{c}{Merged Dataset} \\
        \cline{2-9}
        AR-MINER & 0.49 & 0.34 & 0.4 & \multicolumn{3}{|c|}{No micro or macro average} & 401.57 & NA \\
        \cline{2-9}
        SUR-MINER & 0.66 & 0.55 & 0.6 & 0.64 & 0.6 & 0.62 & 156.54 & 31.31 \\
        \cline{2-9}
        Ensemble & 0.76 & 0.7 & 0.73 & 0.79 & \textbf{0.76} & 0.77 & 176.89 & 29.48 \\
        \cline{2-9}
        DL+WE & \textbf{0.84} & \textbf{0.78} & \textbf{0.81} & \textbf{0.82} & 0.74 & \textbf{0.78} & 155.66 & \textbf{25.94} \\
        \hline
    \end{tabular}
    }
    \end{center}
    \label{table:RQ1_prior}
    \end{table}
    
\begin{table}[htbp]
    \caption{RQ1 results for PTMs. The best scores for each dataset are shown in bold.}
    \begin{center}
    \renewcommand{\arraystretch}{1.2}
    \resizebox{1.0\linewidth}{!}{%
    \begin{tabular}{l | c | c | c | c | c | c | c | c | c}
        \hline
	    {} & \multicolumn{3}{c|}{Micro-avg Scores} & \multicolumn{3}{c|}{Macro-avg scores} & \multirow{3}{4em}{Training Time (s)} & \multirow{3}{4em}{Prediction Time (s)}  & \multirow{3}{4em}{Time-diff (\%)}\\
	    \cline{2-7}
        {} & P & R & F1 & P & R & F1 & {} & {} \\
        \cline{1-7}
        \textbf{Dataset-1} & \multicolumn{6}{c|}{} & & \\
        \cline{2-10}								
        BERT & 0.79 & 0.74 & 0.764 & 0.76 & 0.69 & 0.723 & 345.8 & 19.21 & 34.34\\
        \cline{2-10}
        ALBERT & \textbf{0.93} & \textbf{0.91} & \textbf{0.92} & \textbf{0.89} & \textbf{0.84} & \textbf{0.864} & 286.5 & \textbf{15.92} & 11.33\\
        \cline{2-10}
        RoBERTa & 0.89 & 0.82 & 0.854 & 0.85 & 0.82 & 0.835 & 357.35 & 19.85 & 38.81\\
        \cline{2-10}
        XLNET & 0.76 & 0.68 & 0.718 & 0.72 & 0.68 & 0.699 & 514.92 & 36.78 & 157.2\\
        \cline{2-10}
        \hline
        
        \textbf{Dataset-2} & \multicolumn{8}{c}{}\\
        \cline{2-10}									
        BERT & 0.77 & 0.69 & 0.728 & 0.86 & 0.83 & 0.845 & 185.14 & 10.29 & 165.21\\
        \cline{2-10}
        ALBERT & \textbf{0.96} & \textbf{0.88} & \textbf{0.918} & \textbf{0.94} & \textbf{0.89} & \textbf{0.914} & 119.93 & \textbf{6.66} & 71.65\\
        \cline{2-10}
        RoBERTa & 0.9 & 0.85 & 0.874 & 0.79 & 0.74 & 0.764 & 202.44 & 11.25 & 189.95\\
        \cline{2-10}
        XLNET & 0.85 & 0.77 & 0.808 & 0.87 & 0.78 & 0.823 & 345.57 & 24.68 & 536.08\\
        \cline{2-10}
        \hline
        
        \textbf{Dataset-3} & \multicolumn{8}{c}{}\\
        \cline{2-10}									
        BERT & \textbf{0.93} & \textbf{0.91} & \textbf{0.92} & \textbf{0.9} & \textbf{0.82} & \textbf{0.858} & 156.9 & 6.82 & 111.8\\
        \cline{2-10}
        ALBERT & 0.91 & 0.87 & 0.89 & 0.89 & 0.79 & 0.837 & 105.53 & \textbf{4.59} & 42.55\\
        \cline{2-10}
        RoBERTa & \textbf{0.93} & 0.88 & 0.904 & 0.89 & 0.81 & 0.848 & 175.29 & 7.62 & 136.65\\
        \cline{2-10}
        XLNET & 0.88 & 0.86 & 0.87 & 0.79 & 0.73 & 0.759 & 307.66 & 17.09 & 430.75\\
        \cline{2-10}
        \hline
        
        \textbf{Dataset-4} & \multicolumn{8}{c}{}\\
        \cline{2-10}									
        BERT & 0.84 & 0.78 & 0.809 & 0.79 & 0.72 & 0.753 & 75.69 & 2.91 & 41.95\\
        \cline{2-10}
        ALBERT & 0.85 & 0.81 & 0.83 & 0.72 & 0.65 & 0.683 & 53.66 & \textbf{2.06} & 0.49\\
        \cline{2-10}
        RoBERTa & \textbf{0.93} & \textbf{0.9} & \textbf{0.915} & \textbf{0.87} & \textbf{0.81} & \textbf{0.839} & 87.61 & 3.37 & 64.39\\
        \cline{2-10}
        XLNET & 0.78 & 0.75 & 0.765 & 0.8 & 0.76 & 0.779 & 149.36 & 7.11 & 246.83\\
        \cline{2-10}
        \hline
        
        \textbf{Dataset-5} & \multicolumn{8}{c}{}\\
        \cline{2-10}									
        BERT & 0.88 & 0.81 & 0.844 & 0.8 & 0.77 & 0.785 & 74.52 & 2.57 & 97.69\\
        \cline{2-10}
        ALBERT & 0.93 & \textbf{0.89} & 0.91 & \textbf{0.93} & \textbf{0.88} & \textbf{0.904} & 38.63 & \textbf{1.33} & 2.31\\
        \cline{2-10}
        RoBERTa & \textbf{0.94} & \textbf{0.89} & \textbf{0.914} & 0.87 & 0.83 & 0.85 & 80.69 & 2.78 & 113.85\\
        \cline{2-10}
        XLNET & 0.87 & 0.79 & 0.828 & 0.924 & 0.85 & \textbf{0.904} & 127.46 & 4.9 & 276.92\\
        \cline{2-10}
        \hline
        
        \textbf{Dataset-6} & \multicolumn{8}{c}{}\\
        \cline{2-10}									
        BERT & 0.79 & 0.74 & 0.764 & 0.84 & 0.82 & 0.83 & 74.83 & 2.77 & 92.36\\
        \cline{2-10}
        ALBERT & 0.8 & \textbf{0.77} & 0.785 & \textbf{0.93} & 0.86 & 0.894 & 48.74 & \textbf{1.81} & 25.69\\
        \cline{2-10}
        RoBERTa & \textbf{0.82} & \textbf{0.77} & \textbf{0.794} & \textbf{0.93} & \textbf{0.91} & \textbf{0.92} & 81.2 & 3.01 & 109.03\\
        \cline{2-10}
        XLNET & 0.73 & 0.66 & 0.693 & \textbf{0.93} & 0.85 & 0.888 & 143.15 & 6.22 & 331.94\\
        \cline{2-10}
        \hline
        
        \textbf{Dataset-7} & \multicolumn{8}{c}{Merged Dataset} \\
        \cline{2-10}						
        BERT & 0.91 & 0.84 & 0.874 & 0.88 & 0.8 & 0.844 & 574.32 & 24.97 & -3.74\\
        \cline{2-10}
        ALBERT & 0.95 & 0.85 & 0.897 & 0.85 & 0.8 & 0.824 & 460.19 & \textbf{20.01} & -22.86\\
        \cline{2-10}
        RoBERTa & \textbf{0.97} & \textbf{0.91} & \textbf{0.939} & 0.87 & 0.82 & 0.838 & 597.65 & 25.98 & 0.15\\
        \cline{2-10}
        XLNET & 0.92 & 0.86 & 0.889 & \textbf{0.89} & \textbf{0.84} & \textbf{0.864} & 836.96 & 49.23 & 89.78\\
        \cline{2-10}
        \hline
    \end{tabular}
    }
    \end{center}
    \label{table:RQ1_ptm}
    \end{table}
    
\begin{table}[htbp]
    \caption{RQ2 results for C-PTM-X1 (PTM-X1 = ALBERT). The best scores for each dataset are bold.}
    \begin{center}
    \renewcommand{\arraystretch}{1.5}
    \resizebox{1.0\linewidth}{!}{%
    \begin{tabular}{l | c | c | c | c | c | c | c | c | c}
        \hline
	    {} & \multicolumn{3}{c|}{Micro-avg Scores} & \multicolumn{3}{c|}{Macro-avg scores} & \multirow{3}{4em}{Training Time (s)} & \multirow{3}{4em}{Prediction Time (s)} & \multirow{3}{4em}{Time-Diff (\%)}\\
	    \cline{2-7}
        {} & P & R & F1 & P & R & F1 & {} & {} \\
        \cline{1-7}
        \textbf{Dataset-1} & \multicolumn{6}{c|}{} & & \\
        \cline{2-10}								
        ALBERT (PTM-X1)      &   0.93    &	0.91    &	0.92    &	0.89    &	0.84    &	0.864   &	286.5   &	15.92   &   11.33 \\
        \cline{2-10}
        C-ALBERT-BASE    &   0.94	&   0.916	&   0.928	&   0.902	&   0.844	&   0.872	&   295.1   &   15.6    &   9.09 \\
        \cline{2-10}
        C-ALBERT-MEDIUM  &   0.95	&   0.926	&   0.938	&   0.908	&   0.865	&   0.886	&   280.77  &   \textbf{15.44}  &   7.97 \\
        \cline{2-10}
        C-ALBERT-LARGE   &   \textbf{0.977}	&   \textbf{0.945}	&   \textbf{0.961}	&   \textbf{0.92}	&   \textbf{0.87}	&   \textbf{0.894}	&   295.1   &   15.6    &   9.09 \\
        \cline{2-10}
        \hline
        
        \textbf{Dataset-2} & \multicolumn{8}{c}{}\\
        \cline{2-10}									
        ALBERT (PTM-X1)      &   0.96	&   0.88	&   0.918	&   0.94	&   0.89	&   0.914	&   119.93  &   \textbf{6.66}   &   71.65 \\
        \cline{2-10}
        C-ALBERT-BASE    &   0.972	&   0.888	&   0.928	&   0.947	&   0.894	&   0.92	&   117.53  &   6.79    &   75 \\
        \cline{2-10}
        C-ALBERT-MEDIUM  &   0.987	&   0.906	&   0.945	&   0.958	&   0.903	&   0.93	&   123.53  &   6.86    &   76.8 \\
        \cline{2-10}
        C-ALBERT-LARGE   &   \textbf{0.991}	&   \textbf{0.91}	&   \textbf{0.955}	&   \textbf{0.978}	&   \textbf{0.933}	&   \textbf{0.955}	&   121.13  &   6.86    &   76.8 \\
        \cline{2-10}
        \hline
        
        \textbf{Dataset-3} & \multicolumn{8}{c}{}\\
        \cline{2-10}									
        ALBERT (PTM-X1)      &   0.91    &	0.87    &	0.886   &	0.8     &	0.75    &	0.774   &	105.53  &	4.59    &   42.55 \\
        \cline{2-10}
        C-ALBERT-BASE    &   0.938	&   0.922	&   0.93	&   0.91	&   0.829	&   0.868	&   107.64  &   4.64    &   44.1 \\
        \cline{2-10}
        C-ALBERT-MEDIUM  &   0.949	&   0.926	&   0.937	&   0.914	&   0.841	&   0.876	&   106.59  &   \textbf{4.5}    &   39.75 \\
        \cline{2-10}
        C-ALBERT-LARGE   &   \textbf{0.965}	&   \textbf{0.945}	&   \textbf{0.955}	&   \textbf{0.941}	&   \textbf{0.845}	&   \textbf{0.89}	&   104.47  &   4.64    &   44.1 \\
        \cline{2-10}
        \hline
        
        \textbf{Dataset-4} & \multicolumn{8}{c}{}\\
        \cline{2-10}									
        ALBERT (PTM-X1)      &   0.85    &	0.81    &	0.83    &	0.72    &	0.65    &	0.683   &	53.66   &	2.06    &   0.49 \\
        \cline{2-10}
        C-ALBERT-BASE    &   0.94	&   0.907	&   0.923	&   0.875	&   0.821	&   0.847	&   54.2    &   \textbf{2.02}   &   -1.46 \\
        \cline{2-10}
        C-ALBERT-MEDIUM  &   0.945	&   0.92	&   0.932	&   0.886	&   0.832	&   0.858	&   53.12   &   2.04    &   -0.49 \\
        \cline{2-10}
        C-ALBERT-LARGE   &   \textbf{0.976}	&   \textbf{0.945}	&   \textbf{0.96}	&   \textbf{0.9}	    &   \textbf{0.841}	&   \textbf{0.87}	&   52.59   &   2.08    &   1.46 \\
        \cline{2-10}
        \hline
        
        \textbf{Dataset-5} & \multicolumn{8}{c}{}\\
        \cline{2-10}									
        ALBERT (PTM-X1)      &   0.93    &	0.89    &	0.91    &	0.93    &	0.88    &	0.904   &	38.63   &	1.33    &   2.31 \\
        \cline{2-10}
        C-ALBERT-BASE    &   0.952	&   0.899	&   0.925	&   0.942	&   0.892	&   0.916	&   39.4    &   \textbf{1.3}    &   0 \\
        \cline{2-10}
        C-ALBERT-MEDIUM  &   0.965	&   0.917	&   0.94	&   0.957	&   0.898	&   0.927	&   37.86   &   \textbf{1.3}    &   0 \\
        \cline{2-10}
        C-ALBERT-LARGE   &   \textbf{0.984}	&   \textbf{0.922}	&   \textbf{0.952}	&   \textbf{0.967}	&   \textbf{0.908}	&   \textbf{0.937}	&   38.24   &   1.34    &   3.08 \\
        \cline{2-10}
        \hline
        
        \textbf{Dataset-6} & \multicolumn{8}{c}{}\\
        \cline{2-10}									
        ALBERT (PTM-X1)      &   0.8     &	0.77    &	0.785   &	0.93    &	0.86    &	0.894   &	48.74   &	1.81    &   25.69 \\
        \cline{2-10}
        C-ALBERT-BASE    &   0.831	&   0.775	&   0.802	&   0.938	&   0.921	&   0.929	&   49.71   &   1.83    &   27.08 \\
        \cline{2-10}
        C-ALBERT-MEDIUM  &   0.842	&   0.791	&   0.816	&   0.95	&   0.924	&   0.937	&   49.71   &   \textbf{1.77}   &   22.92 \\
        \cline{2-10}
        C-ALBERT-LARGE   &   \textbf{0.86}	&   \textbf{0.799}	&   \textbf{0.828}	&   \textbf{0.963}	&   \textbf{0.943}	&   \textbf{0.953}	&   49.71   &   1.83    &   27.08 \\
        \cline{2-10}
        \hline
        
        \textbf{Dataset-7} & \multicolumn{8}{c}{Merged Dataset} \\
        \cline{2-10}						
        ALBERT (PTM-X1)     &   0.95    &	0.85    &	0.897   &	0.85    &	0.8     &   0.824   &	460.19  &	20.01   &   -22.86 \\
        \cline{2-10}
        C-ALBERT-BASE    &   0.98	&   0.92	&   0.949	&   0.903	&   0.848	&   0.875	&   464.79  &   20.21   &   -22.09 \\
        \cline{2-10}
        C-ALBERT-MEDIUM  &   0.993	&   0.933	&   0.962	&   0.908	&   0.86	&   0.883	&   450.99  &   20.41   &   -21.32 \\
        \cline{2-10}
        C-ALBERT-LARGE   &   \textbf{0.998}	&   \textbf{0.942}	&   \textbf{0.979}	&   \textbf{0.935}	&   \textbf{0.866}	&   \textbf{0.899}	&   469.39  &   \textbf{19.81}  &   -23.63 \\
        \cline{2-10}
        \hline
    \end{tabular}
    }
    \end{center}
    \label{table:RQ2_cptm_albert}
    \end{table}
    
\begin{table}[htbp]
    \caption{RQ2 results for C-PTM-X2 (PTM-X2 = RoBERTa). The best scores are bold.}
    \begin{center}
    \renewcommand{\arraystretch}{1.5}
    \resizebox{1.0\linewidth}{!}{%
    \begin{tabular}{l | c | c | c | c | c | c | c | c | c}
        \hline
	    {} & \multicolumn{3}{c|}{Micro-avg Scores} & \multicolumn{3}{c|}{Macro-avg scores} & \multirow{3}{4em}{Training Time (s)} & \multirow{3}{4em}{Prediction Time (s)} & \multirow{3}{4em}{Time Diff (\%)}\\
	    \cline{2-7}
        {} & P & R & F1 & P & R & F1 & {} & {} \\
        \cline{1-7}
        \textbf{Dataset-1} & \multicolumn{6}{c|}{} & & \\
        \cline{2-10}								
        RoBERTa (PTM-X2)	    &   0.89    &	0.82    &	0.854   &	0.85    &	0.82    &	0.835   &	357.35  &	19.85   &   38.81 \\
        \cline{2-10}
        C-RoBERTa-BASE	&   0.949	&   0.925	&   0.937	&   0.903	&   0.85	&   0.876	&   360.92	&   19.65   &   37.41 \\
        \cline{2-10}
        C-RoBERTa-MEDIUM	&   0.958	&   0.959	&   0.958	&   0.933	&   0.869	&   0.9	    &   360.92	&   20.25   &   41.61 \\
        \cline{2-10}
        C-RoBERTa-LARGE	&   \textbf{0.991}	&   \textbf{0.982}	&   \textbf{0.996}	&   \textbf{0.97}	&   \textbf{0.911}	&   \textbf{0.94}	&   364.5	&   \textbf{19.45}   &   36.01 \\
        \cline{2-10}
        \hline
        
        \textbf{Dataset-2} & \multicolumn{8}{c}{}\\
        \cline{2-10}									
        RoBERTa (PTM-X2)	    &   0.9     &	0.85    &	0.874   &	0.79    &	0.74    &	0.764   &	202.44  &	11.25   &   189.95 \\
        \cline{2-10}
        C-RoBERTa-BASE	&   0.985	&   0.895	&   0.938	&   0.949	&   0.902	&   0.925	&   204.46	&   11.36   &   192.78 \\
        \cline{2-10}
        C-RoBERTa-MEDIUM	&   0.962	&   0.914	&   0.937	&   0.989	&   0.938	&   0.963	&   200.42	&   \textbf{11.03}   &   184.28 \\
        \cline{2-10}
        C-RoBERTa-LARGE	&   \textbf{0.985}	&  \textbf{ 0.956}	&   \textbf{0.97}	&   \textbf{0.997}	&   \textbf{0.946}	&   \textbf{0.971}	&   204.46	&   11.36   &   192.78 \\
        \cline{2-10}
        \hline
        
        \textbf{Dataset-3} & \multicolumn{8}{c}{}\\
        \cline{2-10}									
        RoBERTa (PTM-X2)	    &   0.93    &	0.88    &	0.904   &	0.89    &	0.81    &	0.848   &	175.29  &	7.62    &   136.65 \\
        \cline{2-10}
        C-RoBERTa-BASE	&   0.953	&   0.931	&   0.942	&   0.925	&   0.845	&   0.883	&   173.54	&   7.7 &   139.13 \\
        \cline{2-10}
        C-RoBERTa-MEDIUM	&   0.978	&   0.942	&   0.96	&   0.927	&   0.859	&   0.892	&   177.04	&   \textbf{7.47}    &   131.99 \\
        \cline{2-10}
        C-RoBERTa-LARGE	&   \textbf{0.995}	&   \textbf{0.986}	&   \textbf{0.99}	&   \textbf{0.965}	&   \textbf{0.884}	&   \textbf{0.923}	&   177.04	&   \textbf{7.47}    &   131.99 \\
        \cline{2-10}
        \hline
        
        \textbf{Dataset-4} & \multicolumn{8}{c}{}\\
        \cline{2-10}									
        RoBERTa (PTM-X2)	    &   0.93	&   0.9	    &   0.915	&   0.87	&   0.81	&   0.839	&   87.61	&   \textbf{3.37}    &   64.39 \\
        \cline{2-10}
        C-RoBERTa-BASE	&   0.943	&   0.912	&   0.927	&   0.879	&   0.82	&   0.848	&   88.49	&   3.44    &   67.8 \\
        \cline{2-10}
        C-RoBERTa-MEDIUM	&   0.966	&   0.94	&   0.953	&   0.897	&   0.841	&   0.868	&   89.36	&   3.4 &   65.85 \\
        \cline{2-10}
        C-RoBERTa-LARGE	&   \textbf{0.986}	&   \textbf{0.962}	&   \textbf{0.974}	&   \textbf{0.947}	&   \textbf{0.877}	&   \textbf{0.911}	&   89.36	&   3.4 &   65.85 \\
        \cline{2-10}
        \hline
        
        \textbf{Dataset-5} & \multicolumn{8}{c}{}\\
        \cline{2-10}									
        RoBERTa (PTM-X2)	    &   0.94	&   0.89	&   0.914	&   0.87	&   0.83	&   0.85	&   80.69	&   2.78    &   113.85 \\
        \cline{2-10}
        C-RoBERTa-BASE	&   0.95	&   0.917	&   0.933	&   0.948	&   0.901	&   0.924	&   79.08	&   2.84    &   118.46 \\
        \cline{2-10}
        C-RoBERTa-MEDIUM	&   0.982	&   0.917	&   0.948	&   0.974	&   0.917	&   0.945	&   81.5	&   \textbf{2.75}    &   111.54 \\
        \cline{2-10}
        C-RoBERTa-LARGE	&   \textbf{0.997}	&   \textbf{0.949}	&   \textbf{0.972}	&   \textbf{0.974}	&   \textbf{0.947}	&   \textbf{0.96}	&   79.08	&   2.81    &   116.15 \\
        \cline{2-10}
        \hline
        
        \textbf{Dataset-6} & \multicolumn{8}{c}{}\\
        \cline{2-10}									
        RoBERTa (PTM-X2)	    &   0.82	&   0.77	&   0.794	&   0.93	&   0.91	&   0.92	&   81.2	&   3.01    &   109.03 \\
        \cline{2-10}
        C-RoBERTa-BASE	&   0.833	&   0.78	&   0.806	&   0.954	&   0.925	&   0.939	&   82.01	&   3.07    &   113.19 \\
        \cline{2-10}
        C-RoBERTa-MEDIUM	&   0.869	&   0.802	&   0.834	&   0.961	&   0.945	&   0.953	&   79.58	&   \textbf{2.95}    &   104.86 \\
        \cline{2-10}
        C-RoBERTa-LARGE	&   \textbf{0.874}	&   \textbf{0.837}	&   \textbf{0.855}	&   \textbf{0.987}	&   \textbf{0.97}	&   \textbf{0.978}	&   82.82	&   2.98    &   106.94 \\
        \cline{2-10}
        \hline
        
        \textbf{Dataset-7} & \multicolumn{8}{c}{Merged Dataset} \\
        \cline{2-10}						
        RoBERTa (PTM-X2)	    &   0.97	&   0.91	&   0.939	&   0.87	&   0.82	&   0.838	&   597.65	&   25.98   &   0.15 \\
        \cline{2-10}
        C-RoBERTa-BASE	&   0.98	&   0.921	&   0.95	&   0.903	&   0.856	&   0.879	&   585.7	&   \textbf{25.46}   &   -1.85 \\
        \cline{2-10}
        C-RoBERTa-MEDIUM	&   0.983	&   0.938	&   0.96	&   0.926	&   0.889	&   0.907	&   603.63	&   25.72   &   -0.85 \\
        \cline{2-10}
        C-RoBERTa-LARGE	&   \textbf{0.995}	&   \textbf{0.975}	&   \textbf{0.984}	&   \textbf{0.949}	&   \textbf{0.895}	&   \textbf{0.921}	&   591.67	&   26.5    &   2.16 \\
        \cline{2-10}
        \hline
    \end{tabular}
    }
    \end{center}
    \label{table:RQ2_cptm_roberta}
    \end{table}

\begin{table}[htbp]
    \caption{RQ3 results for Binary Classifications (Part i)}
    \begin{center}
    \renewcommand{\arraystretch}{1.2}
    \resizebox{1.0\linewidth}{!}{%
    \begin{tabular}{l | c | c | c | c | c}
        \hline
        {\textbf{Datasets} \& Approaches} & P & R & F1 & {Training Time (s)} & {Prediction Time (s)} \\
        \hline
        \textbf{Dataset-1} & \multicolumn{5}{c}{Randomly Selected Class: Bug Report}\\	
        \hline			
        AR-MINER    &   0.37    &   0.36    &   0.36    &   260.92  &   NA \\
        \cline{2-6}
        SUR-MINER   &   0.58    &   0.56    &   0.57    &   105.63  &   21.08 \\
        \cline{2-6}
        Ensemble    &   0.73    &   0.72    &   0.72    &   115.61  &   23.05 \\
        \cline{2-6}
        DL+WE   &   0.87    &   0.73    &   0.79    &   \textbf{99.72}   &   \textbf{14.3} \\
        \cline{2-6}
        BERT    &   0.81    &   0.75    &   0.78    &   344.42  &   19.29 \\
        \cline{2-6}
        ALBERT  &   0.95    &   0.92    &   0.93    &   287.93  &   15.95 \\
        \cline{2-6}
        RoBERTa &   0.91    &   0.83    &   0.87    &   358.78  &   19.87 \\
        \cline{2-6}
        XLNET   &   0.78    &   0.69    &   0.73    &   514.41  &   36.74 \\
        \cline{2-6}
        C-ALBERT-BASE    &   0.95    &   0.95    &   0.95    &   293.92  &   15.66 \\
        \cline{2-6}
        C-ALBERT-MEDIUM  &   0.96    &   0.95    &   0.95    &   281.89  &   15.52 \\
        \cline{2-6}
        C-ALBERT-LARGE   &   0.97    &   0.98    &   0.97    &   295.99  &   15.62 \\
        \cline{2-6}
        C-RoBERTa-BASE   &   0.98    &   0.95    &   0.96    &   361.64  &   19.55 \\
        \cline{2-6}
        C-RoBERTa-MEDIUM &   0.98    &   0.97    &   0.97    &   359.84  &   20.17 \\
        \cline{2-6}
        C-RoBERTa-LARGE  &   \textbf{1}   &   \textbf{0.99}    &   \textbf{0.99}   &   366.32  &   19.37 \\
        \hline
        \textbf{Dataset-2} & \multicolumn{5}{c}{Randomly Selected Class: Inquiry}\\
        \hline
        AR-MINER    &   0.5 &   0.32    &   0.39 &   80.93   &   NA \\
        \cline{2-6}
        SUR-MINER   &   0.6 &   0.56    &   0.58    &   27.04   &   \textbf{3.88} \\
        \cline{2-6}
        Ensemble    &   0.81    &   0.7 &   0.75    &   29.78   &   4.99 \\
        \cline{2-6}
        DL+WE   &   0.94    &   0.86    &   0.9 &   \textbf{26.34}   &   5.29 \\
        \cline{2-6}
        BERT    &   0.8 &   0.7 &   0.75    &   184.58  &   10.32 \\
        \cline{2-6}
        ALBERT  &   0.98    &   0.91    &   0.94    &   120.29  &   6.69 \\
        \cline{2-6}
        RoBERTa &   0.93    &   0.86    &   0.89    &   203.45  &   11.21 \\
        \cline{2-6}
        XLNET   &   0.87    &   0.78    &   0.82    &   345.92  &   24.66 \\
        \cline{2-6}
        C-ALBERT-BASE    &   0.91    &   0.94    &   0.92    &   117.77  &   6.8 \\
        \cline{2-6}
        C-ALBERT-MEDIUM  &   0.92    &   0.95    &   0.93    &   124.15  &   6.87 \\
        \cline{2-6}
        C-ALBERT-LARGE   &   0.92    &   0.98    &   0.95    &   120.65  &   6.87 \\
        \cline{2-6}
        C-RoBERTa-BASE   &   \textbf{1}   &   0.92    &   0.96    &   203.85  &   11.39 \\
        \cline{2-6}
        C-RoBERTa-MEDIUM &   0.98    &   0.93    &   0.95    &   200.02  &   11.07 \\
        \cline{2-6}
        C-RoBERTa-LARGE  &   \textbf{1}   &   \textbf{0.99}    &   \textbf{0.99}   &   204.66  &   11.33 \\
        \hline
        \textbf{Dataset-3} & \multicolumn{5}{c}{Randomly Selected Class: Reliability}\\		
        \hline
        AR-MINER    &   0.44    &   0.36    &   0.4    &   74.77   &   NA \\
        \cline{2-6}
        SUR-MINER   &   0.63    &   0.65    &   0.64    &   \textbf{19.43}   &   \textbf{3.21} \\
        \cline{2-6}
        Ensemble    &   0.79    &   0.77    &   0.78    &   21.42   &   4.29 \\
        \cline{2-6}
        DL+WE   &   0.87    &   0.86    &   0.86    &   19.76   &   3.97 \\
        \cline{2-6}
        BERT    &   0.93    &   0.93    &   0.93    &   157.53  &   6.83 \\
        \cline{2-6}
        ALBERT  &   0.93    &   0.88    &   0.9 &   105.95  &   4.59 \\
        \cline{2-6}
        RoBERTa &   0.95    &   0.89    &   0.92    &   175.64  &   7.66 \\
        \cline{2-6}
        XLNET   &   0.89    &   0.89    &   0.89    &   307.04  &   17.16 \\
        \cline{2-6}
        C-ALBERT-BASE    &   0.94    &   0.95    &   0.92    &   107.1   &   4.65 \\
        \cline{2-6}
        C-ALBERT-MEDIUM  &   0.93    &   0.94    &   0.95    &   107.02  &   4.49 \\
        \cline{2-6}
        C-ALBERT-LARGE   &   0.96    &   0.96    &   0.96    &   104.68  &   4.63 \\
        \cline{2-6}
        C-RoBERTa-BASE   &   0.98    &   0.97    &   0.96    &   173.37  &   7.72 \\
        \cline{2-6}
        C-RoBERTa-MEDIUM &   \textbf{1}   &   0.97    &   0.98    &   177.93  &   7.5 \\
        \cline{2-6}
        C-RoBERTa-LARGE  &   \textbf{1}   &   \textbf{1}   &   \textbf{1.0}   &   177.22  &   7.46 \\
        \hline
        \textbf{Dataset-4} & \multicolumn{5}{c}{Randomly Selected Class: Feature Request}\\	
        \hline					
        AR-MINER    &   0.46    &   0.34    &   0.39    &   71.67   &   NA \\
        \cline{2-6}
        SUR-MINER   &   0.64    &   0.59    &   0.61 &   \textbf{11.76}   &   2.35 \\
        \cline{2-6}
        Ensemble    &   0.85    &   0.75    &   0.8 &   13.54   &   2.7 \\
        \cline{2-6}
        DL+WE   &   0.88    &   0.78    &   0.83    &   12.23   &   {2.05} \\
        \cline{2-6}
        BERT    &   0.86    &   0.8 &   0.83    &   75.84   &   2.9 \\
        \cline{2-6}
        ALBERT  &   0.86    &   0.83    &   0.84    &   53.93   &   2.06 \\
        \cline{2-6}
        RoBERTa &   0.96    &   0.93    &   0.94    &   87.87   &   3.36 \\
        \cline{2-6}
        XLNET   &   0.79    &   0.78    &   0.78    &   148.91  &   7.14 \\
        \cline{2-6}
        C-ALBERT-BASE    &   0.94    &   0.94    &   0.9 &   54.15   &   \textbf{2.02} \\
        \cline{2-6}
        C-ALBERT-MEDIUM  &   0.94    &   0.94    &   0.91    &   53.33   &   2.03 \\
        \cline{2-6}
        C-ALBERT-LARGE   &   0.96    &   0.97    &   0.91    &   52.38   &   2.07 \\
        \cline{2-6}
        C-RoBERTa-BASE   &   0.97    &   0.95    &   0.94    &   88.31   &   3.42 \\
        \cline{2-6}
        C-RoBERTa-MEDIUM &   \textbf{1}   &   0.97    &   0.98    &   89.27   &   3.39 \\
        \cline{2-6}
        C-RoBERTa-LARGE  &   \textbf{1}   &   \textbf{0.99}    &   \textbf{0.99}   &   89  &   3.42 \\
        \hline
    \end{tabular}
    }
    \end{center}
    \label{table:RQ3_binary_a}
    \end{table}

\begin{table}[htbp]
    \caption{RQ3 results for Binary Classifications (Part ii)}
    \begin{center}
    \renewcommand{\arraystretch}{1.5}
    \resizebox{1.0\linewidth}{!}{%
    \begin{tabular}{l | c | c | c | c | c}
        \hline
        {\textbf{Datasets} \& Approaches} & P & R & F1 & {Training Time (s)} & {Prediction Time (s)} \\
        \hline
        \textbf{Dataset-5} & \multicolumn{5}{c}{Randomly Selected Class: App Problem}\\	
        \hline				
        AR-MINER    &   0.54    &   0.35    &   0.42    &   56.96   &   NA \\
        \cline{2-6}
        SUR-MINER   &   0.63    &   0.56    &   0.59    &   \textbf{9.08}    &   1.3 \\
        \cline{2-6}
        Ensemble    &   0.88    &   0.77    &   0.82    &   10.94   &   2.18 \\
        \cline{2-6}
        DL+WE   &   0.85    &   0.76    &   0.8    &   9.72    &   1.62 \\
        \cline{2-6}
        BERT    &   0.91    &   0.82    &   0.86    &   74.45   &   2.58 \\
        \cline{2-6}
        ALBERT  &   0.95    &   0.91    &   0.93    &   38.82   &   1.32 \\
        \cline{2-6}
        RoBERTa &   0.97    &   0.92    &   0.94    &   80.53   &   2.78 \\
        \cline{2-6}
        XLNET   &   0.9 &   0.81    &   0.85    &   127.08  &   4.92 \\
        \cline{2-6}
        C-ALBERT-BASE    &   0.91    &   0.94    &   0.92    &   39.6    &   \textbf{1.29} \\
        \cline{2-6}
        C-ALBERT-MEDIUM  &   0.93    &   0.97    &   0.95    &   37.82   &   1.3 \\
        \cline{2-6}
        C-ALBERT-LARGE   &   0.94    &   0.98    &   0.96    &   38.35   &   1.35 \\
        \cline{2-6}
        C-RoBERTa-BASE   &   0.98    &   0.92    &   0.95    &   79.48    &   2.83 \\
        \cline{2-6}
        C-RoBERTa-MEDIUM &   \textbf{1}   &   0.93    &   0.96    &   81.34   &   2.74 \\
        \cline{2-6}
        C-RoBERTa-LARGE  &   \textbf{1}   &   \textbf{0.97}    &   \textbf{0.98}    &   79.48   &   2.82 \\
        \hline
        \textbf{Dataset-6} & \multicolumn{5}{c}{Randomly Selected Class: Bug Report}\\	
        \hline			
        AR-MINER    &   0.48    &   0.42    &   0.45    &   71.98   &   NA \\
        \cline{2-6}
        SUR-MINER   &   0.59    &   0.53    &   0.56    &   \textbf{10.02}   &   \textbf{1.44} \\
        \cline{2-6}
        Ensemble    &   0.76    &   0.75    &   0.75    &   11.92   &   1.72 \\
        \cline{2-6}
        DL+WE   &   0.78    &   0.77    &   0.77    &   10.32   &   1.47 \\
        \cline{2-6}
        BERT    &   0.81    &   0.75    &   0.78    &   75.13   &   2.76 \\
        \cline{2-6}
        ALBERT  &   0.82    &   0.79    &   0.8 &   48.79   &   1.82 \\
        \cline{2-6}
        RoBERTa &   0.84    &   0.78    &   0.81    &   81.52   &   3.01 \\
        \cline{2-6}
        XLNET   &   0.75    &   0.67    &   0.71    &   142.72  &   6.24 \\
        \cline{2-6}
        C-ALBERT-BASE    &   0.79    &   0.81    &   0.8    &   49.96   &   1.83 \\
        \cline{2-6}
        C-ALBERT-MEDIUM  &   0.81    &   0.84    &   0.82    &   49.91   &   1.78 \\
        \cline{2-6}
        C-ALBERT-LARGE   &   0.82    &   \textbf{0.86}    &   0.84    &   49.51   &   1.82 \\
        \cline{2-6}
        C-RoBERTa-BASE   &   0.85    &   0.79    &   0.82    &   82.17   &   3.08 \\
        \cline{2-6}
        C-RoBERTa-MEDIUM &   0.88    &   0.81    &   0.84    &   79.66   &   2.94 \\
        \cline{2-6}
        C-RoBERTa-LARGE  &   \textbf{0.89}    &   0.84    &   \textbf{0.86}    &   82.65   &   2.99 \\
        \hline
        \textbf{Dataset-7} & \multicolumn{5}{c}{Merged Dataset; Randomly Selected Class: Aspect Evaluation}\\		
        \hline			
        AR-MINER    &   0.51    &   0.35    &   0.42    &   402.77  &   NA \\
        \cline{2-6}
        SUR-MINER   &   0.68    &   0.56    &   0.61    &   157.01  &   31.25 \\
        \cline{2-6}
        Ensemble    &   0.78    &   0.71    &   0.74    &   177.6   &   29.45 \\
        \cline{2-6}
        DL+WE   &   0.86    &   0.79    &   0.82    &   \textbf{156.13}  &   26.04 \\
        \cline{2-6}
        BERT    &   0.93    &   0.86    &   0.89    &   573.75  &   25.04 \\
        \cline{2-6}
        ALBERT  &   0.97    &   0.87    &   0.92    &   462.49  &   20.05 \\
        \cline{2-6}
        RoBERTa &   \textbf{1.0}   &   0.93    &   0.96    &   600.64  &   25.88 \\
        \cline{2-6}
        XLNET   &   0.94    &   0.88    &   0.91    &   834.45  &   48.98 \\
        \cline{2-6}
        C-ALBERT-BASE    &   0.94    &   0.98    &   0.96    &   463.4   &   20.11 \\
        \cline{2-6}
        C-ALBERT-MEDIUM  &   0.96    &   0.99    &   0.97    &   450.09  &   20.51 \\
        \cline{2-6}
        C-ALBERT-LARGE   &   0.96    &   \textbf{1.0}   &   0.98  &   467.98  &   \textbf{19.79} \\
        \cline{2-6}
        C-RoBERTa-BASE   &   0.99    &   0.93    &   0.96    &   584.53  &   25.33 \\
        \cline{2-6}
        C-RoBERTa-MEDIUM &   0.99    &   0.95    &   0.97    &   605.44  &   25.8 \\
        \cline{2-6}
        C-RoBERTa-LARGE  &   \textbf{1.0}   &   0.98    &   \textbf{0.99}   &   592.26  &   26.47 \\
        \hline
    \end{tabular}
    }
    \end{center}
    \label{table:RQ3_binary_b}
    \end{table}
    
\begin{table}[htbp]
    \caption{RQ3 results for Zero-Shot Classification}
    \begin{center}
    \renewcommand{\arraystretch}{1.1}
    \resizebox{1.0\linewidth}{!}{%
    \begin{tabular}{l | c | c | c | c}
        \hline
        {Dataset \#} & P & R & F1 & {Prediction Time (s)} \\
        \hline
        
        \textbf{Dataset-1} & \multicolumn{4}{c}{}\\
        \cline{2-5}					
        AR-MINER    &   0.36    &   0.36    &   0.36    &   N.A. \\
        \cline{2-5}
        ALBERT-NLI  &   0.47    &   0.42    &   0.44    &   \textbf{15.35} \\
        \cline{2-5}
        RoBERTa-NLI &   0.5 &   0.46    &   0.48    &   20.35 \\
        \cline{2-5}
        C-ALBERT-LARGE-NLI   &   0.5 &   0.47    &   0.48    &   16.7 \\
        \cline{2-5}
        C-RoBERTa-LARGE-NLI  &   0.59    &   0.53    &   0.56    &   20.66 \\
        \cline{2-5}
        RoBERTa-LARGE-MNLI  &   \textbf{0.64}    &   \textbf{0.6}   &   \textbf{0.62}    &   19.1 \\
        \hline
        
        \textbf{Dataset-2} & \multicolumn{4}{c}{}\\
        \cline{2-5}					
        AR-MINER    &   0.49    &   0.32    &   0.39    &   N.A. \\
        \cline{2-5}
        ALBERT-NLI  &   0.46    &   0.42    &   0.44    &   6.98 \\
        \cline{2-5}
        RoBERTa-NLI &   0.5 &   0.47    &   0.48    &   10.85 \\
        \cline{2-5}
        C-ALBERT-LARGE-NLI   &   0.51    &   0.48    &   0.49    &   \textbf{6.47} \\
        \cline{2-5}
        C-RoBERTa-LARGE-NLI  &   0.56    &   0.5 &   0.53    &   11.64 \\
        \cline{2-5}
        RoBERTa-LARGE-MNLI  &   \textbf{0.64}    &   \textbf{0.58}    &  \textbf{0.61}    &   10.8 \\
        \hline
        
        \textbf{Dataset-3} & \multicolumn{4}{c}{}\\
        \cline{2-5}					
        AR-MINER    &   0.44    &   0.36    &   0.4 &   N.A. \\
        \cline{2-5}
        ALBERT-NLI  &   0.43    &   0.4 &   0.41    &   \textbf{4.37} \\
        \cline{2-5}
        RoBERTa-NLI &   0.51    &   0.46    &   0.48    &   7.28 \\
        \cline{2-5}
        C-ALBERT-LARGE-NLI   &   0.55    &   0.51    &   0.53    &   4.72 \\
        \cline{2-5}
        C-RoBERTa-LARGE-NLI  &   0.61    &   0.57    &   0.59    &   7.95 \\
        \cline{2-5}
        RoBERTa-LARGE-MNLI  &   \textbf{0.71}    &   \textbf{0.65}    &   \textbf{0.68}    &   7.78 \\
        \hline
        
        \textbf{Dataset-4} & \multicolumn{4}{c}{}\\
        \cline{2-5}
        AR-MINER    &   0.45    &   0.33    &   0.38    &   N.A. \\
        \cline{2-5}
        ALBERT-NLI  &   0.42    &   0.38    &   0.4 &   \textbf{2.00} \\
        \cline{2-5}
        RoBERTa-NLI &   0.51    &   0.47    &   0.49    &   3.22 \\
        \cline{2-5}
        C-ALBERT-LARGE-NLI   &   0.47    &   0.42    &   0.44    &   2.12 \\
        \cline{2-5}
        C-RoBERTa-LARGE-NLI  &   0.65    &   0.59    &   0.62    &   3.51 \\
        \cline{2-5}
        RoBERTa-LARGE-MNLI  &   \textbf{0.69}    &   \textbf{0.64}    &   \textbf{0.66}    &   3.21 \\
        \hline
        
        \textbf{Dataset-5} & \multicolumn{4}{c}{}\\
        \cline{2-5}					
        AR-MINER    &   0.53    &   0.35    &   0.42    &   N.A. \\
        \cline{2-5}
        ALBERT-NLI  &   0.39    &   0.35    &   0.37    &   \textbf{1.37} \\
        \cline{2-5}
        RoBERTa-NLI &   0.53    &   0.49    &   0.51    &   2.91 \\
        \cline{2-5}
        C-ALBERT-LARGE-NLI   &   0.47    &   0.44    &   0.45    &   \textbf{1.37} \\
        \cline{2-5}
        C-RoBERTa-LARGE-NLI  &   0.59    &   0.54    &   0.56    &   2.72 \\
        \cline{2-5}
        RoBERTa-LARGE-MNLI  &   \textbf{0.74}    &   \textbf{0.67}    &   \textbf{0.70} &   2.9 \\
        \hline
        
        \textbf{Dataset-6} & \multicolumn{4}{c}{}\\
        \cline{2-5}				
        AR-MINER    &   0.47    &   0.41    &   0.44    &   N.A. \\
        \cline{2-5}
        ALBERT-NLI  &   0.4 &   0.36    &   0.38    &   \textbf{1.73} \\
        \cline{2-5}
        RoBERTa-NLI &   0.43    &   0.4 &   0.41    &   2.95 \\
        \cline{2-5}
        C-ALBERT-LARGE-NLI   &   0.43    &   0.41    &   0.42    &   1.88 \\
        \cline{2-5}
        C-RoBERTa-LARGE-NLI  &   0.52    &   0.48    &   0.5 &   3.13 \\
        \cline{2-5}
        RoBERTa-LARGE-MNLI  &   \textbf{0.64}    &   \textbf{0.58}    &   \textbf{0.61}    &   2.87 \\
        \hline
        
        \textbf{Dataset-7} & \multicolumn{4}{c}{Merged Dataset} \\
        \cline{2-5}			
        AR-MINER    &   0.49    &   0.34    &   0.4 &   N.A. \\
        \cline{2-5}
        ALBERT-NLI  &   0.47    &   0.44    &   0.45    &   \textbf{19.11} \\
        \cline{2-5}
        RoBERTa-NLI &   0.58    &   0.53    &   0.55    &   24.94 \\
        \cline{2-5}
        C-ALBERT-LARGE-NLI   &   0.48    &   0.44    &   0.46    &   19.39 \\
        \cline{2-5}
        C-RoBERTa-LARGE-NLI  &   0.65    &   0.61    &   0.63    &   24.84 \\
        \cline{2-5}
        RoBERTa-LARGE-MNLI  &   \textbf{0.74}    &   \textbf{0.68}    &   \textbf{0.71}    &   27.28 \\
        \hline
    \end{tabular}
    }
    \end{center}
    \label{table:RQ3_zero_shot}
    \end{table}
    
\begin{table}[htbp]
    \caption{RQ3 results for Multi-Task (Category) Classification}
    \begin{center}
    \renewcommand{\arraystretch}{1.2}
    \resizebox{1.0\linewidth}{!}{%
    \begin{tabular}{l | c | c | c | c | c | c | c | c}
        \hline
	    {} & \multicolumn{3}{c|}{Micro-Scores} & \multicolumn{3}{c|}{Macro-Scores} &    \multirow{2}{4em}{Training Time (s)}  &   \multirow{2}{4em}{Prediction Time (s)} \\
	    \cline{2-7}
        {} & P & R & F1 & P & R & F1 & {} & {}\\
        \hline
        Category Labeled Data & \multicolumn{8}{c}{}  \\
        \cline{2-9}
        \hline
        BERT    &   0.813	&   0.759	&   0.791	&   0.752	&   0.714	&   0.732	&   287.16  &   9.988   \\
        \cline{2-9}
        ALBERT  &   0.883	&   0.865	&   0.86	&   0.842	&   0.785	&   0.823	&   230.095 &   12.006  \\
        \cline{2-8}
        RoBERTa &   0.844	&   0.782	&   0.81	&   0.736	&   0.693	&   0.698	&   239.06  &   10.392  \\
        \cline{2-9}
        XLNET   &   0.798	&   0.717	&   0.753	&   0.822	&   0.721	&   0.779	&   418.48  &   19.692  \\
        \cline{2-9}
        \hline	
        C-ALBERT-BASE    &  0.888	&   0.85	&   0.872	&   0.853	&   0.785	&   0.807	&   \textbf{139.437} &   12.126  \\
        \cline{2-9}
        C-ALBERT-MEDIUM  &  0.897	&   0.864	&   0.897	&   0.865	&   0.809	&   0.837	&   180.396 &   6.123   \\
        \cline{2-9}
        C-ALBERT-LARGE   &  \textbf{0.933}	&   0.89	&   \textbf{0.917}	&   0.859	&   0.804	&   0.831	&   281.634 &   \textbf{5.943}   \\
        \cline{2-9}
        \hline	
        C-RoBERTa-BASE   &  0.945	&   0.837	&   0.878	&   0.887	&   0.858	&   0.86	&   292.85  &   10.184  \\
        \cline{2-9}
        C-RoBERTa-MEDIUM &  0.894	&   0.86	&   0.877	&   \textbf{0.942}	&   0.894	&   0.904	&   301.815 &   12.86   \\
        \cline{2-9}
        C-RoBERTa-LARGE  &  0.932	&   \textbf{0.90}	&   0.901	&   0.931	&   \textbf{0.879}	&   \textbf{0.927}	&   295.835 &   10.6    \\
        \hline
        
    \end{tabular}
    }
    \end{center}
    \label{table:RQ3_multi_task}
    \end{table}

    \begin{table}[htbp]
        \caption{RQ3 results for Multi-Task (Sentiment) Classification}
        \begin{center}
        \renewcommand{\arraystretch}{1.2}
        \resizebox{1.0\linewidth}{!}{%
        \begin{tabular}{l | c | c | c | c | c | c | c | c}
            \hline
            {} & \multicolumn{3}{c|}{Micro-Scores} & \multicolumn{3}{c|}{Macro-Scores} &    \multirow{2}{4em}{Training Time (s)}  &   \multirow{2}{4em}{Prediction Time (s)} \\
            \cline{2-7}
            {} & P & R & F1 & P & R & F1 & {} & {}\\
            \hline
            Sentiment Labeled Data & \multicolumn{8}{c}{}  \\
            \cline{2-9}
            \hline
            BERT    &   0.852	&   0.769	&   0.804	&   0.759	&   0.748	&   0.739	&   689.184	    &   32.461  \\
            \cline{2-9}
            ALBERT  &   0.886	&   0.889	&   0.877	&   0.848	&   0.806	&   0.822	&   552.228	    &   26.013  \\
            \cline{2-9}
            RoBERTa &   0.862	&   0.805	&   0.83	&   0.75	&   0.711	&   0.74	&   836.71	    &   31.176  \\
            \cline{2-9}
            XLNET   &   0.808	&   0.737	&   0.787	&   0.849	&   0.754	&   0.794	&   1088.048	&   63.999  \\
            \cline{2-9}
            \hline	
            C-ALBERT-BASE    &  0.908	&   0.883	&   0.884	&   0.868	&   0.796	&   0.838	&   604.227	    &   26.273  \\
            \cline{2-9}
            C-ALBERT-MEDIUM  &  0.913	&   0.878	&   0.89	&   0.879	&   0.829	&   0.845	&   \textbf{541.188}	    &   26.533  \\
            \cline{2-9}
            C-ALBERT-LARGE   &  0.93	&   0.911	&   0.918	&   0.892	&   0.823	&   0.858	&   610.207	    &   \textbf{23.772}  \\
            \cline{2-9}
            \hline	
            C-RoBERTa-BASE   &  0.958	&   0.861	&   0.904	&   0.913	&   0.858	&   0.903	&   644.27	    &   33.098  \\
            \cline{2-9}
            C-RoBERTa-MEDIUM &  0.934	&   0.864	&   0.913	&   0.955	&   0.905	&   0.929	&   784.719	    &   28.292  \\
            \cline{2-9}
            \textbf{C-RoBERTa-LARGE}  &  \textbf{0.964}	&   \textbf{0.934}	&   \textbf{0.93}	&   \textbf{0.972}	&   \textbf{0.907}	&   \textbf{0.935}	&   828.338	    &   31.800  \\
            \hline
            
        \end{tabular}
        }
        \end{center}
        \label{table:RQ3_multi_task_sentiment}
        \end{table}

\begin{table}[htbp]
    \caption{RQ3 results for Multi-Resource Classification (Trained on App Reviews and Tested on Twitter)}
    \begin{center}
    \renewcommand{\arraystretch}{1.2}
    \resizebox{0.75\linewidth}{!}{%
    \begin{tabular}{l | c | c | c | c }
        \hline
        {\textbf{Datasets} \& Approaches} & P & R & F1 & {Prediction Time (s)} \\
        \hline
        \textbf{Dataset-2} & \multicolumn{4}{c}{}\\	
        \hline				
        AR-Miner    &   0.42	&   0.28	&   0.34    &   N.A. \\
        \cline{2-5}				
        SUR-Miner    &   0.61	&   0.56	&   0.58    &   39.22 \\
        \cline{2-5}				
        Ensemble    &   0.78	&   0.72	&   0.75    &   38.79 \\
        \cline{2-5}				
        DL+WE    &   0.85	&   0.78	&   0.81    &   27.79 \\
        \cline{2-5}
        \hline
        BERT    &   0.57	&   0.50	&   0.53    &   17.49 \\
        \cline{2-5}
        ALBERT  &   0.69	&   0.65	&   0.67    &   \textbf{9.99} \\
        \cline{2-5}
        RoBERTa &   0.67	&   0.60	&   0.63    &   18.0 \\
        \cline{2-5}
        XLNET   &   0.64	&   0.56	&   0.60    &   44.42 \\
        \cline{2-5}
        \hline	
        C-ALBERT-BASE    &   0.75	&   0.67	&   0.71    &   11.54 \\
        \cline{2-5}
        C-ALBERT-MEDIUM  &   0.79	&   0.75	&   0.77    &   10.29 \\
        \cline{2-5}
        C-ALBERT-LARGE   &   0.84	&   0.76	&   0.80    &   12.35 \\
        \cline{2-5}
        \hline	
        C-RoBERTa-BASE   &   0.80	&   0.74	&   0.77    &   20.45 \\
        \cline{2-5}
        C-RoBERTa-MEDIUM &   0.82	&   0.77	&   0.79    &   19.85 \\
        \cline{2-5}
        \textbf{C-RoBERTa-LARGE}  &   \textbf{0.87}	&   \textbf{0.86}	&   \textbf{0.86}    &   20.45 \\
        \hline
        
    \end{tabular}
    }
    \end{center}
    \label{table:RQ3_multi_resource}
    \end{table}
    
\begin{table}[htbp]
    \caption{RQ3 results for Multi-Resource Classification (Trained on Twitter and Tested on App Reviews)}
    \begin{center}
    \renewcommand{\arraystretch}{1.2}
    \resizebox{0.75\linewidth}{!}{%
    \begin{tabular}{l | c | c | c | c }
        \hline
        {Datasets \& Approaches} & P & R & F1 & {Prediction Time (s)} \\
        \hline
        Dataset-2 & \multicolumn{4}{c}{}\\	
        \hline				
        AR-Miner    &   0.36	&   0.24	&   0.29	&   N.A \\
        \cline{2-5}				
        SUR-Miner    &   0.66	&   0.59	&   0.62	&   25.1    \\
        \cline{2-5}				
        Ensemble    &   0.83	&   0.76	&   0.79	&   23.66   \\
        \cline{2-5}				
        DL+WE    &   0.89	&   0.83	&   0.86	&   16.4    \\
        \cline{2-5}
        \hline
        BERT    &   0.62	&   0.55	&   0.58	&   9.62    \\
        \cline{2-5}
        ALBERT  &   0.73	&   0.72	&   0.72	&   6.29    \\
        \cline{2-5}
        RoBERTa &   0.70	&   0.64	&   0.67	&   11.7    \\
        \cline{2-5}
        XLNET   &   0.69	&   0.62	&   0.65	&   28.43   \\
        \cline{2-5}
        \hline	
        C-ALBERT-BASE    &   0.79	&   0.70	&   0.74	&   6.81    \\
        \cline{2-5}
        C-ALBERT-MEDIUM  &   0.85	&   0.83	&   0.84	&   5.87    \\
        \cline{2-5}
        C-ALBERT-LARGE   &   0.93	&   0.82	&   0.87	&   6.79    \\
        \cline{2-5}
        \hline	
        C-RoBERTa-BASE   &   0.82	&   0.75	&   0.78	&   13.09   \\
        \cline{2-5}
        C-RoBERTa-MEDIUM &   0.89	&   0.82	&   0.85	&   11.12   \\
        \cline{2-5}
        \textbf{C-RoBERTa-LARGE}  &   \textbf{0.92}	&   \textbf{0.92}	&   \textbf{0.92}	&   11.45   \\
        \hline
        
    \end{tabular}
    }
    \end{center}
    \label{table:RQ3_multi_resource_2}
    \end{table}

\end{document}